\documentclass[preprint]{elsarticle}

\usepackage{hyperref}

\journal{Artificial Intelligence Journal}

 \usepackage{amsmath}
 \usepackage{amsthm}
 \usepackage[utf8]{inputenc}\DeclareUnicodeCharacter{2212}{-}
 \usepackage{pgf}
 \usepackage{tikz}
 \usepackage{amssymb}
 \usepackage{proof_macros}
 \usepackage{algorithm}
 \usepackage[noend]{algpseudocode}
 \usepackage{caption}
 \usepackage{subcaption}
 \usepackage{ifthen}
 \usepackage{multicol}
 \usepackage{float}
 \usepackage{pict2e}
 \graphicspath{ {figures/} }
 \usetikzlibrary{shadings}
 \usetikzlibrary{patterns}

\begin{document}

\begin{frontmatter}
 
\title{X*: Anytime Multi-Agent Path Finding for Sparse Domains using Window-Based Iterative
  Repairs
}


\author{Kyle Vedder}
\ead{kvedder@seas.upenn.edu}

\address{University of Pennsylvania, Department of Computer and Information Science \\ 
  220 South 33rd Street, Philadelphia, PA 19104, United States of America}

\author{Joydeep Biswas}
\ead{joydeepb@cs.utexas.edu}

\address{University of Texas Austin, Department of Computer Science \\ 
2317 Speedway, Austin, TX 78712, United States of America\vspace{-3.1em}}
\begin{abstract}
Real-world multi-agent systems such as warehouse robots operate under
significant time constraints -- in such settings, rather than spending significant amounts of time solving
for \emph{optimal} paths, it is instead preferable to find valid, collision-free
paths quickly, even if suboptimal, and given additional time, to iteratively
refine such paths to improve their cost. In such domains, we observe that
agent-agent collisions are \emph{sparse} -- they involve small local subsets of
agents, and are geographically contained within a small region of the overall
space. 
Leveraging this insight, we can first plan paths for each agent individually, and in the
cases of collisions between agents, perform small local repairs limited to
local subspace \emph{windows}. As time permits, these windows can be successively
grown and the repairs within them refined, thereby improving the path quality,
and eventually converging to the global joint optimal solution.
Using these insights, we present two algorithmic contributions: 1) the Windowed
Anytime Multiagent Planning Framework (\ampp{}) for a class of anytime
planners that quickly generate valid paths with suboptimality estimates and generate optimal paths given sufficient time, and 2) X*, an
efficient \ampp{}-based planner. X* is able to efficiently find
successive valid solutions by employing re-use techniques during the repair
growth step of \ampp{}. 
Experimentally, we demonstrate that in sparse domains: 1) X* outperforms
state-of-the-art anytime or optimal \mapf{} solvers in time to valid path, 2) X*
is competitive with state-of-the-art anytime or optimal \mapf{} solvers in time
to optimal path, 3) X* quickly converges to very tight suboptimality bounds, and
4) X* is competitive with state-of-the-art suboptimal  \mapf{} solvers in time
to valid path for small numbers of agents while providing much higher quality
paths. 

\end{abstract}


\end{frontmatter}

\section{Introduction}\sectionlabel{Introduction}

Multi-Agent Path Finding (\mapf{}) is the problem of finding a collision free, mimimal cost global path $\pth$ in the joint space of the set of agents $\agentset$ traveling from a set of start states $\start$ to a set of goal states $\goal$ on a graph, often with one or more graph edges blocked at runtime~\cite{mapfdefinitions}. The path cost, denoted $\pthcost{\pth}$, is often defined as the makespan of $\pth$ (i.e.\ the maximum cost for any agent) or the sum of costs for each agent; in this work we focus on optimizing for sum of costs, but this choice is not fundamental.  Much of the prior art in \mapf{} focuses on finding optimal or bounded suboptimal global paths for large numbers of densely packed agents, often with a focus on how planners scale with an increasing number of agents~\cite{mapfdefinitions, silvercooperativepathfinding, subdimexp, conflictbasedsearch, Cohen2018AnytimeFS, pushandrotate}; however, there are many real-world multi-agent scenarios that have sparse agent distributions, are highly dynamic, and require valid paths in milliseconds such as warehouse robots~\cite{kivasystems}, robot soccer~\cite{minutebotstdp2017, minutebotstdp2018, mrltdp2019, nubottdp2019}, or drone swarms~\cite{swarmsurvery, robotflyanddrive}. In such scenarios, finding a optimal global path is too time consuming; instead, it is desirable to employ an anytime solver that can quickly find a collision-free global path of reasonable quality and, if given additional time, improve the global path quality, ultimately converging to an optimal global path.

In this work we focus on the problem of producing an anytime planner which, in \emph{sparse} domains, quickly finds a valid global path of reasonable quality and, if given sufficient time, will converge to an optimal global path. As part of this work, we leverage three key insights. 1) Unlike in domains like 8-puzzle~\cite{russellandnorvig} with each tile treated as an independent agent, in \emph{sparse} domains, problem instances often have agent-agent collisions for individually planned global paths that involve only a small subset of the total agents and are isolated to a small area easily separable from other collisions. By exploiting sparsity, the \mapf{} problem can be decomposed into small subspaces, (i.e.\ small subsets of states and agents) and each subspace efficiently searched to produce a \emph{repair} to the collision (i.e.\ a new, collision-free section of the global path for the colliding agents), thus producing a valid global path. 2) These subspaces can trade repair generation time for repair quality by varying their size; growing the area of a subspace will produce a repair of the global path of the same or better \emph{quality} (i.e.\ lower contribution to the global path cost), but takes more time to produce a repair. 3) Iteratively growing the subspace and generating repairs monotonically improves the global path quality. When a repair search proceeds \emph{unimpeded}, i.e.\ unrestricted by the constraints the subspace imposes on the full space, from the global start to the global goal of the agents involved, the global path is known to be optimal for those agents. 

By combining these key insights, we present an anytime \mapf{} framework called \amppfull{} (\ampp{}), along with an efficient \ampp{}-based planner called Expanding A* (X*) that performs search reuse for efficient iterative path repair. Experimentally, we demonstrate that in sparse domains: 

\begin{enumerate}
  \item X* outperforms state-of-the-art anytime or optimal \mapf{} solvers in time to valid path.
  \item X* is competitive with state-of-the-art anytime or optimal \mapf{} solvers in time to optimal path.
  \item X* quickly converges to very tight suboptimality bounds.
  \item X* is competitive with state-of-the-art suboptimal  \mapf{} solvers in time to valid path for small numbers of agents while providing much higher quality paths.
\end{enumerate}

An earlier version of this work presented a similar version of \ampp{}, the na\"ive \ampp{} implementation, and X*~\cite{Vedder2019AAMAS}, but this work provides refined pseudocode, more detailed explanations, walked through examples, and a completely new experimental results section.

The rest of this paper proceeds as follows: We first introduce relevant background (\sectionref{Background}) and provide an overview of related \mapf{} solvers (\sectionref{RelatedWork}). We then present \ampp{}, our \mapf{} solving framework, along with a na\"ive implementation and two worked out examples (\sectionref{\ampp}). We then present X*, an efficient \ampp{}-based planner that performs search reuse for efficient successive path repair (\sectionref{xstar}). Finally, we present several experiments to characterize X* and compare it to prior art in sparse domains (\sectionref{results}), and then discuss directions for future work (\sectionref{FutureWork}).

\section{Background}\sectionlabel{Background}

To put our contributions in the context of the state-of-the-art, we begin by discussing the complexity of Single-Agent Path Finding along with the variety of solution approaches seen in the literature (\subsectionref{sapfbackground}). We then discuss the complexity of Multi-Agent Path Finding, comparing it to the single agent version, along with the variety of solution approaches seen in the literature (\subsectionref{mapfbackground}). We then discuss the breadth of both \sapf{} and \mapf{} prior art that employ three techniques which are relevant to our contributions, namely 
Bounded Search (\subsectionref{BoundedSearch}),
Search Reuse (\subsectionref{SearchReuse}), and
Anytime Path Planning (\subsectionref{AnytimePlanning}). This presentation will prepare the reader for \sectionref{RelatedWork} where we analyze several \mapf{} solvers that utilize these techniques.

\newcommand{\backgroundsubsectionspacing}{\vspace{0em}}


\backgroundsubsectionspacing
\subsection{Single-Agent Path Finding}\subsectionlabel{sapfbackground}

Constructing a minimal cost, collision free path from a known start state to a known goal state for a single agent in the face of obstacles and under time constraints is a problem faced in many domains, from robotics to videogame agents. This problem, known as the Single-Agent Path Finding problem (\sapf{}), appears in domains with both discrete and continuous state spaces.

In discrete spaces, the problem can be modeled in a variety of ways, including 
integer linear programming~\cite{integerlinearprogrammingsapf1,integerlinearprogrammingsapf2}, 
satisfiability~\cite{satsapf}, and answer set 
programming~\cite{answersetprogrammingsapfmapf}; 
however, solutions most commonly model the problem as a graph with vertices that
represent a state in the state space and with edges that represent the valid transitions between
these states. Graph search algorithms are then used to find minimal
 cost paths between the start
vertex and the goal vertex on the graph, and the resulting path can be mapped to a minimal cost set of transitions from the start state to the goal state. These graph search algorithms can be \emph{uninformed}, meaning they know nothing about the problem beyond the given graph (e.g.\ Uniform Cost Search~\cite{russellandnorvig}) or they can be \emph{informed}, meaning they have additional information about the graph such as a heuristic, e.g.\ A*~\cite{astar}, or regular problem structure, e.g.\ Jump Point Search~\cite{jumppointsearch}. 

In continuous spaces, the most computationally challenging problems are intractable; for
linked polyhedra moving through three-dimensional space with a fixed set of
polyhedral obstacles, commonly known as the Moving Sofa problem or the Couch
Mover's problem, finding an optimal, collision free
path is PSPACE hard~\cite{generalizedmoversproblem}. A common way to simplify
continuous problems is to convert them to discrete problems~\cite{dynamicnavmesh,probablisticrobotics}; 
this is often done by imposing a
grid-structure, such as a four-connected grid or an eight-connected grid~\cite{kivasystems,coursetofine}, or by
randomly sampling the space~\cite{prmonchip}. Imposing a grid adds additional structure to the problem
that can be exploited to speed search~\cite{jumppointsearch}, but environments
can be adversarially designed to admit no collision free path along a given grid, but
admit many collision free paths in the continuous space version of the problem.
To address this problem, the search space can be sampled online, ensuring
probabilistic completeness~\cite{samplingplannersurvey}. Two common ways this
is done is by constructing a random graph and then searching it~\cite{prm}
or by constructing the data structure during
search~\cite{rrtstar}; the latter approach enables planning to be joined with perception, thereby
dramatically lowering their overall computational cost~\cite{jointperceptionplanning}.

\backgroundsubsectionspacing
\subsection{Multi-Agent Path Finding}\subsectionlabel{mapfbackground}

The problem of finding collision-free
paths for \emph{multiple} agents that also avoid agent-agent collisions, known as the Multi-Agent Path Finding problem
(\mapf{}), presents another layer of difficulty. Not only is the continuous, two
dimensional case of path finding for multiple rectangles, a simplification of the Couch
Mover's problem setup, PSPACE hard~\cite{pspacehard}, the discrete \mapf{} problem is also
significantly more challenging than the discrete \sapf{} problem. In general, planning jointly for all
agents requires planning in a state space with the dimensionality that is at
least linear in the number of agents, meaning the cardinality of the state space is
at least exponential in the number of agents.
Under common conditions, \sapf{} operates on a \emph{polynomial} domain, i.e.\
the difficulty of the problem grows polynomially relative to the depth of the
optimal solution due to duplicate detection; under
these same conditions, \mapf{} operates on an \emph{exponential}
domain, i.e.\ the difficulty of the problem grows exponentially in the depth of
the solution~\cite{exponentialvspolynomialdomains,exponentialpolynomialdomainspaperintro}. 
Similar to \sapf{}, discrete \mapf{} problems can be modeled via integer linear 
programming~\cite{integerlinearprogrammingmapf}, 
satisfiability~\cite{satmapf,satmapf2,satmapf3}, and answer set 
programming~\cite{answersetprogrammingmapf}, but many solutions operate directly on 
graphs~\cite{silvercooperativepathfinding, subdimexp, Cohen2018AnytimeFS}.

\backgroundsubsectionspacing
\subsection{Bounded Search}\subsectionlabel{BoundedSearch}

Bounded Search is a technique where artificial limits are placed on the search
space. While bounds usually produce a suboptimal solution,
they prevent planning far into the future on a model of the world that is less
likely to be accurate, thereby speeding solution generation. This bound can be
enforced via the time
domain such as with a time-bounded lattice~\cite{Kushleyev2009TimeboundedLF},
via depth of search such as Hierarchical Cooperative
A*~\cite{silvercooperativepathfinding}, or via restricted cost propagation
such as Truncated D* Lite~\cite{truncateddstarlite}.

\backgroundsubsectionspacing
\subsection{Search Reuse}\subsectionlabel{SearchReuse}

Search Reuse is a technique where information from one or more previous searches
is used to
speed up future searches. One of the most widly used families of reuse algorithms, D*~\cite{dstar} / D*
Lite~\cite{dstarlite} and their variants~\cite{truncateddstarlite,anytimedstar,fielddstar}, operates by propagating changes in the environment back up the search
tree, only modifying states \gvalue{}s as needed. Other examples of algorithms
that employ reuse are from the predator-prey domain, where the predator
 prunes the search tree of a prior search to make it suitable for
the current search, thereby saving the cost of re-expanding the remaining states 
in the pruned tree~\cite{fringeretrievingastar,generalizedfringeretrievingastar,movingtargetdstarlite}.

\backgroundsubsectionspacing
\subsection{Anytime Path Planners}\subsectionlabel{AnytimePlanning}

Anytime Path Planners are planners that can quickly develop a solution to the given problem
and, if given more computation time, iteratively improve the plan quality.
Anytime algorithms are desirable for many domains as they allow for
metareasoning to make online tradeoffs between solution quality and planning
time~\cite{metareasoning, metareasoning2, metareasoning3}. A na\"{i}ve way to construct an anytime planner is to run a standard
planner with parameters which trade
solution optimality for a runtime improvement (e.g.\ A* heuristic inflation~\cite{russellandnorvig}),
and then iteratively re-run the planner
with tighter bounds if computation time remains~\cite{anytimeastar}. While this
first plan generation is
often fast, successive iterations grow increasingly slow due to lack of information
reuse. Anytime planners that instead reuse information
from prior searches are typically faster at generating successive
plans~\cite{anytimerepairingastar,anytimewindowastar,anytimemultiheuristicastar}.

There exist other, non A*-like anytime path planners that also leverage reuse
techniques, such as RRT*~\cite{rrtstar}, which finds a feasible solution
and then, given more time, repeatedly improves it by further sampling the space
and updating the tree with cheaper intermediate nodes when applicable,
converging to the optimal solution in the limit. Reuse and bounded search techniques can
also be combined to further speed anytime
search~\cite{anytimetruncateddstar,anytimesipp}.

\section{\mapf{} Related Work}\sectionlabel{RelatedWork}

In this work we focus on \mapf{} solving for general graphs. In principle, any uninformed weighted graph search algorithm such as Uniform Cost Search (UCS)~\cite{russellandnorvig} is sufficient to find an optimal path for any \mapf{} problem by treating each joint state as a position of a single high dimensional meta-agent; however, providing additional information such as a heuristic, framing the problem differently, or exploiting additional properties often present in relevant domains can produce more efficient algorithms, provide different runtime characteristics, or provide different guarantees, thus motivating the variety of \mapf{} solvers.


\mapf{} solvers fall into two major classes: global search and decoupled search. Like UCS, global search techniques solve a single large meta-agent search problem; however, these techniques attempt to leverage problem substructure to speed search~\cite{subdimexp, Ryan2008ExploitingSS,operatordecomposition,Felner2012PartialExpansionAW,enhancedpartialexpansionastar}. Decoupled search approaches decompose the problem by planning for each agent serially, forcing later agents to account for sections or the entirety of earlier agents plans~\cite{silvercooperativepathfinding,conflictbasedsearch,pushandrotate, Erdmann1987,velocitypathdecomposition,Leroy:1999:MPC:1624312.1624378,Saha2006MultiRobotMP,crosbysingleagentapproachmultiagentplanning,suboptimalconflictbasedsearch}. In order to discuss our approach in the context of prior art, we present a unified notation as follows: every state $\state$ is in the joint space of the agent set $\agentset$ which contains one or more possibly heterogeneous agents. In order to refer to the part of $\state$ associated with a subset of its agents, we introduce a state filter function $\filterpath{\state}{\agentset'}$, where $\agentset' \subseteq \agentset$. For example, if $\state$'s agent set $\agentset = \{a, b, c\}$ and we want to refer to the part of $\state$ associated with agents $b$ and $c$, this is denoted by $\filterpath{\state}{\{b, c\}}$. This notation allows us to reason about the subspaces that we introduce shortly. Importantly, in our notation states do not contain time bookkeeping; while time is relevant for collision checking, the bookkeeping for collision checking is well understood~\cite{mapfdefinitions} and abstracted away by the state neighbor function $\neighbors{\state}$, so we omit it for simplicity.

M*~\cite{subdimexp} is a state-of-the-art global \mapf{} solver that exploits domain sparsity in order to speed its search. M* operates by first computing an optimal individual space \emph{policy} to $\filterpath{\goal}{\{ a \} }$ for all $a \in \agentset$. It then traces a path in the space of $\agentset$ from $\start$ to $\goal$ using the policies of each agent. If a collision is encountered, M* is able to use the policy information to compute the relevant $\agentset' \subseteq \agentset$ to involve in a joint search. In sparse domains, the number of agents involved in this joint search is small, allowing M* to avoid the aforementioned combinatorial explosion, and collisions are typically separate from one another, avoiding the need to merge joint searches. Due to the expensive nature of the policy computation for each agent, even if lazily computed with approaches like Reverse Resumable A*~\cite{silvercooperativepathfinding}, M* is ill-suited to the task of quickly generating a valid solution in sparse domains. Furthermore, while M* can produce optimal and $\epsilon$-suboptimal paths, it is not anytime nor does its $\epsilon$-suboptimal version allow for efficient path refinement if given additional time.

Conflict-Based Search (CBS)~\cite{conflictbasedsearch} is a state-of-the-art decoupled \mapf{} solver that exploits domain sparsity to speed search. CBS first computes an optimal path from $\filterpath{\start}{\{ a \} }$ to $\filterpath{\goal}{\{ a \} }$ for all $a \in \agentset$; if a collision occurs between agents $i$ and $j$, CBS forms two models of the world, one where the path of $i$ is constrained through the collision point and the path of $j$ is replanned, and one where the path of $j$ is constrained through the collision point and the path of $i$ is replanned. This approach is then applied recursively to each model, forming a conflict tree. In sparse domains, the number of agents involved in a collision is often small, therefore producing a small conflict tree. A characteristic of CBS is it sometimes struggles with open areas; when there are many short paths that collide and a longer path needs to be employed, the conflict tree grows very large before the optimal solution is considered. Furthermore, while CBS can produce optimal paths and its extended counterpart ECBS can produce $\epsilon$-suboptimal paths~\cite{suboptimalconflictbasedsearch}, neither are anytime nor does ECBS allow for efficient path refinement of $\epsilon$-suboptimal paths if given additional time.

Anytime Focal Search (AFS)~\cite{Cohen2018AnytimeFS} is a state-of-the-art global \mapf{} solver that exploits the availability of ``good enough'' solutions in order to quickly find a valid solution and improves this path if given more time. AFS maintains open set $\openset$ and closed set $\closedset$ structures similar to A* and an additional structure \emph{focal list} of states that have \fvalue{s} of no more than $\epsilon$ times larger than the smallest value in $\openset$. Rather than constraining itself to only expand minimal cost states, AFS is willing to expand other states in the focal list, determined via a \emph{priority function}, thereby allowing it to quickly find a path to $\goal$ that is $\epsilon$-suboptimal. Given more time, the bookkeeping done in the focal list allows AFS to tighten $\epsilon$ and improve its path without searching from scratch, ultimately producing an optimal solution. As AFS is anytime, it is able to provide intermediate results along with a confidence bound. AFS does not attempt to decompose the problem as it always plans in the full joint space of $\agentset$ from $\start$ to $\goal$, leading to higher valid solution runtimes compared to planners that exploit sparsity.

Push and Rotate (PR)~\cite{pushandrotate} is a state-of-the-art decoupled \mapf{} solver. Unlike the other solvers presented, PR does not attempt to find an optimal or bounded suboptimal solution; instead, it uses graph transformations (\texttt{Push} and \texttt{Rotate}) to quickly find a \emph{valid} solution, allowing it to scale to large numbers of agents with highly dense agent distributions. As PR is not an optimal or bounded suboptimal solver, it provides no guarantees of path quality; in our experimentation, PR commonly generated paths of cost 2x greater than optimal paths. Due to the high cost of the generated paths and an inability to refine them, PR is ill-suited for domains that require a high quality path.

Expanding A* (X*), which we introduce, combines many of the strengths of these algorithms. Like CBS, X* first computes an optimal path from $\filterpath{\start}{\{ a \} }$ to $\filterpath{\goal}{\{ a \} }$ for all $a \in \agentset$. Like M*, when a collision is detected, it performs joint search only in a subspace, but without the need to compute individual policies and in a much smaller subspace. Like AFS, X* is able to produce intermediate solutions while also exploiting domain sparsity. Like PR, X* is able to quickly generate a valid solution in sparse domains but with tighter quality bounds.

There exists a number of extensions to CBS and M* that either utilize optimizations to underlying solvers that are orthogonal to the approach itself or exploit regular domain structure when avaiable. Examples of orthogonal optimizations include Operator Decomposition (OD)~\cite{operatordecomposition}, which operates by first considering neighbors that only change the path of one agent; these approaches are applicable to any A*-based solver, including X*. Examples of optimizations that exploit domain structure to speed search include Enhanced Partial Expansion A* (EPEA*)~\cite{enhancedpartialexpansionastar}, which exploits domain structure to only generate a subset of neighbors at a specific \fvalue{} and Prioritize Conflicts in Improved CBS (ICBS)~\cite{improvedconflictbasedsearch}, which relies upon avoiding alternate paths of the same cost for one or both agents involved in pair-wise collisions, as typically found in structured domains, in order to reduce conflict tree size; however, the approaches of X*, AFS, CBS, and M* do not exploit domain structure in this way.

\section{\amppfull}\sectionlabel{\ampp}

As discussed in \sectionref{Introduction}, the size of the joint state space grows exponentially in the number of agents; this motivates subspace-based approaches such as M* that speed up search by decomposing the full MAPF problem into smaller sub-problems involving fewer agents. A key insight is that while subspaces can be used to limit the search to a subset of \emph{agents}, they can also be used to limit the search to a subset of \emph{states}. 

We present a construct called a \emph{window} that encapsulates a subset of agents and a connected subset of states. A window is placed around a collision in the global path in order to produce a \emph{repair} to the global path by performing a search within the window. The start of the repair search in $\windowk$, denoted $\startk$, is the first state on the global path in the window and the goal of the repair search in $\windowk$, denoted $\goalk$, is the last state on the global path in the window. Every window $\windowk$ has a \emph{successor} window $\windowkpone$ that shares the same agent set but has a superset of states. This allows for the concept of iteratively \emph{growing} a window by replacing it with its successor that considers more of the domain in its repair. Two windows can be merged together to form a larger window that incorporates both smaller windows via the $\cup$ operator. For example, $\window$ and $\window'$ can be joined together to form a larger window $\window'' := \window \cup \window'$; $\window''$ must have an agent set $\agentset'' = \agentset \cup \agentset'$ and all of the states in $\window$ and $\window'$ must be part of the joint states of $\window''$. Finally, two windows can be checked for overlap via the $\cap$ operator. For example, $\window \cap \window'$ is true if and only if their agent sets $\agentset$ and $\agentset'$ overlap and they share one or more individual agent states. These window definitions and mechanics are demonstrated  in \subsectionref{amppexamples}.

While a window-based repair does not ensure the resulting repaired global path is optimal, a repair in a successor window $\windowkpone$ ensures that its repaired global path will be \emph{at most} the same cost as the global path repaired by $\windowk$ and often cost less. Thus, repeatedly growing the subspace and generating repairs monotonically improves the global path quality. Furthermore, if a window $\windowk$ is sufficiently large that $\startk$ and $\goalk$ are the global start $\filterpath{\start}{ \agentset }$ and goal $\filterpath{\goal}{ \agentset }$ for its agents $\agentset$ and $\windowk$ does not \emph{impede the search} from $\startk$ to $\goalk$, i.e.\ limit search exploration with $\windowk$ state restrictions, then the joint paths for the agents $\agentset$ in $\windowk$ are jointly optimal and $\windowk$ can be discarded. If no more windows exist, then the joint path is an optimal solution.
Using this insight, we introduce an anytime \mapf{} framework called the \amppfull{} (\ampp). 

\subsection{\ampp{} Overview}

We present the pseudocode for \ampp{} in \algoref{\ampp} featuring the eponymous top level procedure, the recursive procedure \callsmall{\recampp} which does the heavy lifting, and the overlapping window helper \callsmall{\mergecollidingwindows}. The \ampp{} pseudocode only manages the state of search windows; all searches are conducted by the implementation defined components \callsmall{\planin} and \callsmall{\growandreplan}, discussed in \subsectionref{amppcomponents}, in order to make \ampp{} domain-agnostic.

\ampp{} operates by initially forming a potentially colliding global path by planning for each agent in individual space. \callsmall{\recampp} is then invoked, and this recursive procedure makes tail-recursive calls until the global path is provably optimal, each time improving the quality of the global path. \callsmall{\recampp} operates by first growing and replanning in all existing windows, merging them with existing windows if they overlap (\linesref{AMPPGrowAndReplanStart}{AMPPGrowAndReplanEnd}), then creating new windows to encapsulate any remaining collisions, merging them with existing windows if they overlap (\linesref{AMPPWhileExistsCollision}{AMPPEndWhileExistsCollision}). At this point, no more collisions exist in the global path and thus the global path is valid. \callsmall{\recampp} then removes any window searches which have optimally repaired the global path (\linesref{AMPPShouldQuitLoop}{AMPPShouldQuit}); if no more windows exist, then the global path is proven optimal (\lineref{AMPPReportDone}) and \callsmall{\recampp{}} terminates. Otherwise, the current valid global path is reported as an intermediary solution along with its optimality bound estimate. This bound is computed via the current global path cost, an exact or over-estimate of the optimal global path cost, divided by the individual space planned global path cost, an exact estimate or an under-estimate of the optimal global path cost (\lineref{AMPPReportIntermediary}). \callsmall{\recampp{}} then recursively invokes itself for another iteration.

\newcommand{\commentsmall}[1]{}
\begin{algorithm}[htb!]
  \caption{\amppfull}\algolabel{\ampp}
  \begin{algorithmic}[1]
    \Procedure{\ampp}{}
    \State $\pth \gets$ joint plan comprised of optimal paths planned in individual space\linelabel{AMPPPlanIndependently}
    \State $W \gets \emptyset$\linelabel{AMPPWEmpty}
    \State \Return \Call{\recampp}{$\pth, W, \pthcost{\pth}$}
    \EndProcedure

    \Procedure{\recampp}{$\pth, W, \initialpathcost$}\algolabel{\recampp}
    \ForAll {$\windowk \in W$} \linelabel{AMPPGrowAndReplanStart}
    \If {$\exists w' \in W: w' \neq \windowk \land w' \cap \windowkpone$} \linelabel{AMPPCheckInterfere}\commentsmall{If successor overlaps}
    \State $W \gets W \setminus \{ \windowk \} $\linelabel{AMPPRemoveOverlap}
    \State $W, \pth \gets \Call{\mergecollidingwindows}{\windowkpone, W, \pth}$ \linelabel{AMPPOverlapMerge} \commentsmall{Merge}
    \State \Continue
    \EndIf
    \State $\windowkpone, \pth \gets \Call{\growandreplan}{\windowk,
      \pth}$\linelabel{AMPPGrowAndReplan}
    \State $W \gets \left(W \setminus \{ \windowk \} \right) \cup \{ \windowkpone \}$ \linelabel{AMPPGrowAndReplanEnd}
    \EndFor
    \While {$\Call{\firstcollisionwindow}{\pth} \neq \emptyset$}\linelabel{AMPPWhileExistsCollision} \commentsmall{Detect collisions}
    \State $w \gets \Call{\firstcollisionwindow}{\pth}$\linelabel{AMPPMakeWindow}
    \State $W, \pth \gets \Call{\mergecollidingwindows}{w, W,
      \pth}$ \linelabel{AMPPEndWhileExistsCollision} \commentsmall{Merge and plan in}
    \EndWhile
    \ForAll {$w \in W$}\linelabel{AMPPShouldQuitLoop} \commentsmall{Discard windows with globally optimal repairs}
    \If {\Call{\shouldquit}{$\pth, w$}} $W \gets W \setminus \{w\}$     \EndIf \linelabel{AMPPShouldQuit}
    \EndFor
    \If {$W = \emptyset$}
    \Return $(  \pth, 1)$ \linelabel{AMPPReportDone} \commentsmall{Optimal solution found}
    \EndIf
    \State \Report $\left(  \pth, \frac{\pthcost{\pth}}{\initialpathcost} \right)$\linelabel{AMPPReportIntermediary}\commentsmall{Produce intermediary solution and bound}
    \State \Return \Call{\recampp}{$\pth, W, \initialpathcost$}\linelabel{AMPPRecurse}
    \EndProcedure
    
    \Function{\mergecollidingwindows}{$w, W, \pth$}
    \ForAll{$w' \in W:  w' \cap w$}\linelabel{AMPPforallcollidingwindows}
    \State $W \gets W \setminus \{w, w'\}$\linelabel{AMPPRemoveOldWindow}
    \State $w \gets  w \cup w'$\linelabel{AMPPUnionWindows}
    \EndFor
    \State $\pth \gets \Call{\planin}{w, \pth}$\linelabel{AMPPPlanJointly}
    \State $W \gets W \cup \{w\}$\linelabel{AMPPAddWindow}
    \State \Return $ \left( W, \pth  \right)$\linelabel{AMPPMergeCollideReturns}
    \EndFunction
  \end{algorithmic}
\end{algorithm}

One of the important features of \ampp{} is it repairs collisions chronologically, thus ensuring that each window added and repaired is making progress towards a valid path. Newly added and repaired windows can potentially change the relative timing of agents \emph{later} along the path, inadvertently fixing later collisions or introducing new ones; however, these repairs cannot cause changes \emph{earlier} along the path, only later. By sweeping from the beginning to the end of the path, \ampp{} ensures that once a window is added its repair work cannot be undone by other repairs and any collisions induced by a repair must be later along the path and thus handled by \ampp{}.

Another important feature of \ampp{} is it avoids invalidating repair windows during valid path improvement. It is possible that an earlier window can be grown and replanned in, producing a new repair of higher quality that changes the relative time that agents enter a later window; this change would invalidate the start state of the later window, forcing its repair efforts to be discarded. As such, repair searches and improvements are responsible for not invalidating any windows that exist later along the path; this can be implemented via padding as discussed in \subsectionref{nwa} and illustrated in \subsectionref{pathpaddingexample}.

Together, these two features ensure that \ampp{}'s running time for a valid path is a function of the number of agent-agent collisions and their separability from other collisions, i.e.\ domain sparsity, and running time for successive repairs is a function of the number of windows and the number of agents involved in each window.

\subsection{\ampp{} Components}\subsectionlabel{amppcomponents}

  As \ampp{} is a domain agnostic framework for anytime \mapf{} planners, it has several definitions/subroutines which must be provided by any planner implementing it:

\newpage
 \amppcomponent{Window definition:} a window definition is state space specific, but a window $\windowk$ must uphold the aforementioned properties, namely:

 \newcommand{\windowpropertiesspacing}{\vspace{-0.7em}}
 \begin{itemize}
  \windowpropertiesspacing
   \item Contain a connected subset of states for a subset of agents
   \windowpropertiesspacing
   \item Possess a start $\startk$ and a goal $\goalk$ on the global path
   \windowpropertiesspacing
   \item Possess a successor window $\windowkpone$ which contains a superset of states and the same agent set
   \windowpropertiesspacing
   \item The ability to merge with another window to form a new window encapsulating the agent sets and states contained in $\windowk$ and the other window via the $\cup$ operator which returns the new window
   \windowpropertiesspacing
   \item The ability to check for overlap with another window via the $\cap$ operator which returns a boolean
 \end{itemize}

  \amppcomponent{\Call{\firstcollisionwindow}{$\pth$}:} given a path $\pth$, this subroutine finds the first agent-agent collision along the time dimension, beginning with $\pth_0$. If collisions exists, return a window encapsulating the first collision; otherwise, return $\emptyset$.

  \amppcomponent{\Call{\planin}{$\windowk, \pth$}:} the given path $\pth$ has an  associated agent set $\agentset$ and the given window $\windowk$ has an associated agent set $\agentset'$, where $\agentset' \subseteq \agentset$. This subroutine generates a collision free repair in $\windowk$ by planning an optimal path from $\startk$ to $\goalk$, respecting the entry times of agents to $\startk$. The repair is inserted as a replacement to the relevant subset of $\pth$, respecting the relative timings of agents involved in later windows, and $\pth$ is returned. 

  \amppcomponent{\Call{\growandreplan}{$\windowk, \pth$}:} the given path $\pth$ has an associated agent set $\agentset$ and the given window $\windowk$ has an associated agent set $\agentset'$, where $\agentset' \subseteq \agentset$. This subroutine grows $\windowk$ by replacing it with its successor, $\windowkpone$, and generates a repair in $\windowkpone$ by planning an optimal path  from $\startkpone$ to $\goalkpone$, and inserting it as a replacement to the relevant subset of $\pth$, respecting the relative timings of agents involved in later windows, then returning $(\windowkpone, \pth)$. \Call{\growandreplan}{$\windowkpone, \pth$} is guaranteed to only be invoked when \Call{\planin}{$\windowkpone, \pth$} or \Call{\growandreplan}{$\windowk, \pth$} have previously been invoked and is guaranteed that $\windowkpone$ does not overlap with any other existing window.

  \amppcomponent{\Call{\shouldquit}{$\pth, \windowk$}:} this subroutine is a predicate that determines if the given window $\windowk$ should be discarded. In order to ensure that \ampp{} produces globally optimal solutions, a window $\windowk$ with an associated agent set $\agentset$ cannot be discarded until $\startk = \filterpath{\start}{ \agentset }$, $\goalk = \filterpath{\goal}{ \agentset }$, and $\windowk$ does not impede the repair search.


Assuming a \ampp{}-based planner meets these conditions:

  \noindent \textit{Theorem }1. The planner will produce a valid global path after a single iteration of \callsmall{\recampp}{}. See \appendixref{amppproofs}, \theoremref{AMPPValidSolutionImmediately} for proof.

  \noindent \textit{Theorem }2. Given sufficient iterations of \callsmall{\recampp}{}, the planner will produce a optimal global path. See \appendixref{amppproofs}, \theoremref{AMPPOptimal} for proof.

\subsection{Na\"ive Windowing A*}\subsectionlabel{nwa}

\definecolor{darkgreen}{RGB}{0, 100, 0}
\definecolor{orange}{RGB}{255,127,0}
\definecolor{faintgray}{RGB}{200, 200, 200}
\definecolor{wallblack}{RGB}{20, 20, 20}
\newcommand{\windowstyle}{dashed}
\newcommand{\windowwidth}{0.35mm}
\newcommand{\examplefigscale}{0.45}
\newcommand{\gridcolor}{faintgray}
\newcommand{\wallcolor}{wallblack}
\newcommand{\wallwidth}{0.5mm}
\newcommand{\acolor}{red}
\newcommand{\bcolor}{blue}
\newcommand{\ccolor}{darkgreen}
\newcommand{\dcolor}{orange}
\newcommand{\aletter}{\textcolor{\acolor}{a}}
\newcommand{\bletter}{\textcolor{\bcolor}{b}}
\newcommand{\cletter}{\textcolor{\ccolor}{c}}
\newcommand{\dletter}{\textcolor{\dcolor}{d}}
\newcommand{\dimval}{0.2}

To provide a concrete example of a \ampp{}-based planner, we present Na\"ive Windowing A* (NWA*), a na\"ive implementation of \ampp{} with a window definition specific to unit cost four-connected grids. NWA* employs A* as the underlying window solver and makes no attempt at search re-use when the window is grown. We present the requisite \ampp{} definitions/subroutines:

\amppcomponent{Window definition:} The window is formulated as a high dimensional rectangular prism, characterized by its bottom left and upper right corners in the joint space of its agent set. New windows are initialized around a collision state by selecting all states that have an $L_\infty$ distance from the collision state  of less than or equal to a hyperparameter. An example of such a window is shown in \figref{simpleamppwindowrepair1}, where the window, drawn as a dashed rectangle, is in the joint space of $\aletter$ and $\bletter$ and created  via an $L_\infty$ norm of $1$. A window is grown by moving its corners further away from the center by a fixed number of steps. An example of window growth is shown in the transition from \figref{simpleamppwindowrepair1} to \figref{simpleamppwindowrepair2}, where window is grown by increasing the radius by a state. Windows $\window$ and $\window'$ overlap if $\agentset \cap \agentset' \neq \emptyset$ and their rectangles overlap. An example of non-overlapping windows is shown in \figref{amppwindowrepair2}, and an example of overlapping windows is shown in \figref{amppexamplewindowsoverlap}. Windows $\window$ and $\window'$ are merged to create $\window''$ by unioning their agent sets and constructing a containing rectangle. An example of a window merge is shown in \figref{amppexamplewindowmerge}, where $\window^{\aletter \bletter}$ and $\window^{\aletter \cletter}$ merge to form $\window^{\aletter \bletter \cletter}$. 

\amppcomponent{\Call{\firstcollisionwindow}{$\pth$}:} This subroutine looks for collisions along the global path $\pth$, starting with $\pth_0$ and ending with state $\pth_{\len{\pth} - 1}$. If a collision is detected, a window is initialized around the colliding state with the colliding agents; otherwise, $\emptyset$ is returned.

\amppcomponent{\Call{\planin}{$\windowk, \pth$}:} the given global path $\pth$ has an  associated agent set $\agentset$ and the given window $\windowk$ has an associated agent set $\agentset'$, where $\agentset' \subseteq \agentset$. $\startk$ and $\goalk$ are computed from $\filterpath{\pth}{\agentset'}$; $\startk$ is the first state on $\filterpath{\pth}{\agentset'}$ in $\window$ and $\goalk$ is the last state on $\filterpath{\pth}{\agentset'}$ in $\window$. An A* search is run the in the space of $\windowk$ from $\startk$ to $\goalk$, with any expanded state's neighboring states not in $\window$ discarded rather than placed in the open set $\openset$. The resulting repair $\pth'$ replaces the section of path in $\filterpath{\pth}{\agentset'}$ from $\startk$ to $\goalk$. Importantly, if $\pth$ is \emph{not} already a valid solution, then $\pth$'s cost may stay the same or it may increase after $\pth'$ is inserted; if $\pth$ \emph{is} already a valid solution, then $\pth'$ will be of the same or reduced cost compared to the region of $\filterpath{\pth}{\agentset'}$ from $\startk$ to $\goalk$, as $\pth$ will have already been repaired by a window $\windowkmone$, and so the larger $\windowk$ may find a repair $\pth'$ for the same region of $\filterpath{\pth}{\agentset'}$ that costs less. In the case where $\pth'$ costs less, it must be padded in order to ensure all agents leave $\goalk$ at the same time as they did in prior to the insertion of $\pth'$ in $\pth$; an example of this is shown in \subsectionref{pathpaddingexample}. Additionally, if the A* search returns \nopath{}, $\windowk$ is grown to form $\windowkpone$ and the result of \Call{\planin}{$\windowkpone, \pth$} is returned.

\amppcomponent{\Call{\growandreplan}{$\windowk, \pth$}:} This subroutine grows $\windowk$ by replacing it with its successor, $\windowkpone$, and then returning the result of \Call{\planin}{$\windowkpone, \pth$}.

\amppcomponent{\Call{\shouldquit}{$\pth, \windowk$}:} the global path $\pth$ has an associated agent set $\agentset$ and the window $\windowk$ has an associated agent set $\agentset'$. This subroutine returns true iff $\startk$ and $\goalk$ are $\filterpath{\pth}{\agentset'}_0$ and $\filterpath{\pth}{\agentset'}_{\len{\pth} - 1}$, respectively, and $\windowk$ did not impede the search during the last invocation of \Call{\planin}{$\windowk, \pth$}, i.e.\ neighbors were not culled during any of A*'s state expansion due to $\windowk$'s state space constraints.

\subsection{\ampp{} Examples
}\subsectionlabel{amppexamples}

In order to illustrate the behavior of \ampp{} (\algoref{\ampp}), we present four worked out examples. The first example (\figref{amppwalkthroughsimple}) demonstrates how \ampp{} operates for a single collision between two agents using NWA*'s window definition. The second example (\figref{amppwalkthrough}) demonstrates how \ampp{} operates for multiple collisions using NWA*'s window definition. The third example (\figref{ampppaddingexample}) demonstrates how \ampp{} can generate valid but suboptimal solutions, and how path insertion and padding operates using NWA*'s window definition. The fourth example (\figref{ampparbgraph}) demonstrates how \ampp{} can operate on arbitrary graphs and how NWA*'s window definition can be generalized. All examples are applicable to NWA* as well as our efficient \ampp{}-based planner, X* (\sectionref{xstar}), as both planners share the same window definition. The first three examples operate on a $10 \times 10$ unit cost four-connected grid and the fourth example operates on a random graph.

\subsubsection{Single Window Example}

The single window example shown in \figref{amppwalkthroughsimple} demonstrates the mechanics of window creation, window growth and replanning, and window termination using NWA*'s window definition. The example demonstrates a single collision between two agents resolved via a window search; this window is then repeatedly expanded and re-searched until it encompasses an unimpeded search from $\start$ to $\goal$. The associated figures depict how \ampp{} planning for agents individually can induce a collision (\figref{simpleamppexindividualplans}), how a window encapsulates a repair and what a repair looks like for joint plans (\figref{simpleamppwindowrepair1}), how a window can be grown to consider a larger search space, therefore potentially improving repair quality (\figref{simpleamppwindowrepair2}), and that a window can be terminated after it encapsulates a repair from the start to the goal and does not impede the repair search (\figref{simpleamppwindowrepair3}). 
A key takeaway from this example is that \ampp{} windows do not need to encapsulate the entirety of the potentially infinite number of states in the space of their agents in order to terminate.

A line-by-line analysis of \figref{amppwalkthroughsimple} grounded in the \ampp{} algorithm (\algoref{WAMPF}) is as follows:

\noindent
\begin{tabular}{p{25.5em}|p{7.5em}}
  \hline
  \statementjustif{Optimal paths are planned for each agent individually to form a global path; the paths for agents $\aletter$ and $\bletter$ collide at Step 1. }{\linesref{AMPPPlanIndependently}{AMPPWEmpty}. Shown in \figref{simpleamppexindividualplans}.}

  \statementjustif{\callsmall{\recampp{}} invoked. There are no existing windows, so no window manipulations are done. }{\linesref{AMPPGrowAndReplanStart}{AMPPGrowAndReplanEnd}.}

  \statementjustif{The collision between $\aletter$ and $\bletter$ is detected by \callsmall{\firstcollisionwindow} and $\window^{\aletter \bletter}$ is formed to encapsulate it.}{\linesref{AMPPWhileExistsCollision}{AMPPMakeWindow}.}



  \statementjustif{\callsmall{\mergecollidingwindows} is invoked to merge $\window^{\aletter \bletter}$ with existing windows if needed; however, there are no existing windows ($W$ is empty) so no merging occurs.}{\linesref{AMPPforallcollidingwindows}{AMPPRemoveOldWindow}.}

  \statementjustif{\callsmall{\planin} is invoked to generate a repair in $\window^{\aletter \bletter}$. $\window^{\aletter \bletter}$ is added to the window set $W$.}{\linesref{AMPPPlanJointly}{AMPPAddWindow}. Shown in \figref{simpleamppwindowrepair1}.}

  \statementjustif{No more collisions exist so \callsmall{\firstcollisionwindow} returns $\emptyset$ and the collision detection loop exits.}{\lineref{AMPPforallcollidingwindows}.}

  \statementjustif{$\window^{\aletter \bletter}$ does not allow for an unimpeded search from $\filterpath{\start}{\{ \aletter, \bletter \}}$ to $\filterpath{\goal}{\{ \aletter, \bletter \}}$, so \callsmall{\shouldquit} returns false and $W$ remains unchanged.}{\linesref{AMPPShouldQuitLoop}{AMPPShouldQuit}.}

  \statementjustif{$W$ is not empty so the global path $\pth$ is not returned as optimal, but it is reported as an intermediary solution along with its optimality bound.}{\linesref{AMPPReportDone}{AMPPReportIntermediary}.}

  \statementjustif{\callsmall{\recampp{}} is recursively invoked, with $W = \{\window^{\aletter \bletter} \}$ and a valid but potentially suboptimal global path.}{\lineref{AMPPPlanJointly}.}

  \statementjustif{$\window^{\aletter \bletter}$ is grown and replanned in, producing a larger $\window^{\aletter \bletter}$ and a repair. The larger $\window^{\aletter \bletter}$ replaces its predecessor in $W$, and it does not overlap with any other windows so no merging is done. }{\linesref{AMPPGrowAndReplan}{AMPPGrowAndReplanEnd}.}

\end{tabular}

\input{ampp_example_figure_small.tex}

\noindent
\begin{tabular}{p{25.5em}|p{7.5em}}

  \statementjustif{No collisions exist and $\window^{\aletter \bletter}$ does not allow for an unimpeded search from $\filterpath{\start}{\{ \aletter, \bletter \}}$ to $\filterpath{\goal}{\{ \aletter, \bletter \}}$, so the updated global path is reported as an intermediary solution and \callsmall{\recampp} is recursively invoked.}{ \linesref{AMPPWhileExistsCollision}{AMPPReportDone}. Shown in \figref{simpleamppwindowrepair2}.}

  \statementjustif{\callsmall{\recampp} proceeds, growing $\window^{\aletter \bletter}$ and updating its repair and intermediary solutions, with no collisions introduced. The repair in $\window^{\aletter \bletter}$  allowed for an unimpeded search from $\filterpath{\start}{\{ \aletter, \bletter \}}$ to $\filterpath{\goal}{\{ \aletter, \bletter \}}$, therefore allowing \callsmall{\shouldquit} to return true. This removes $\window^{\aletter \bletter }$ from $W$, making $W$ empty and thus returns the global path as optimal. }{\linesref{AMPPGrowAndReplanStart}{AMPPReportDone}. Shown in \figref{simpleamppwindowrepair3}.}
  \hline
\end{tabular}


\subsubsection{Multi-Window Example}

The example shown in \figref{amppwalkthrough} expands on the mechanics demonstrated in \figref{amppwalkthroughsimple} by demonstrating window merging and subspace planning capabilities using NWA*'s window definition. The example demonstrates a collision between two agents whose repair causes a cascading collision with another agent later along the path. The two repairs are then grown, eventually merging into the joint space of three agents, and eventually terminates after allowing an unimpeded search from $\start$ to $\goal$. The associated figures depict how \ampp{} planning for agents individually can induce a collision, but often only for a subset of agents (\figref{amppexindividualplans}), how a window repair can cause collisions later in the path, creating the need for more windows (\figref{amppwindowrepair1}), the creation of a second window, finally generating a collision free solution (\figref{amppwindowrepair2}), that grown windows which overlap in the state and agent space need to be merged (\figref{amppexamplewindowsoverlap}), the resulting merged window (\figref{amppexamplewindowmerge}), and the repeatedly grown window which is finally terminated (\figref{amppexamplewindowmax}).
A key takeaway from this example is \ampp{}'s window-based approach speeds search; while the given problem involves four agents, \ampp{} never required a search in the joint space of more than three agents to produce an optimal path and only required two small searches in the joint space of two agents to produce a valid path.

A line-by-line analysis of \figref{amppwalkthrough} grounded in the \ampp{} algorithm (\algoref{WAMPF}) is as follows:

\noindent
\begin{tabular}{p{25.5em}|p{7.5em}}
  \hline
  \statementjustif{Plans optimal paths for each agent individually and $W$ is initialized. Note that agents $\aletter$ and $\bletter$ collide at step 1. }{\linesref{AMPPPlanIndependently}{AMPPWEmpty}. Shown in \figref{amppexindividualplans}.}

  \statementjustif{\callsmall{\recampp{}} invoked. The collision between $\aletter$ and $\bletter$ is detected by \callsmall{\firstcollisionwindow} and $\window^{\aletter \bletter}$ is formed to encapsulate it, and there are no windows to collide with.}{\linesref{AMPPGrowAndReplanStart}{AMPPRemoveOldWindow}.}

  \statementjustif{\callsmall{\planin} is invoked to generate a repair in $\window^{\aletter \bletter}$. $\window^{\aletter \bletter}$ is added to the window set $W$.}{\linesref{AMPPPlanJointly}{AMPPAddWindow}. Shown in \figref{amppwindowrepair1}.}

  \statementjustif{The $\window^{\aletter \bletter}$ repair has created a new collision later in time between $\aletter$ and $\cletter$. On the next iteration of the loop \callsmall{\firstcollisionwindow} detects the collision and $\window^{\aletter \cletter}$ is formed to encapsulate it.}{\linesref{AMPPWhileExistsCollision}{AMPPMakeWindow}.}

  \statementjustif{\callsmall{\mergecollidingwindows} is invoked to merge $\window^{\aletter \cletter}$ with existing windows as needed, but $W = \{ \window^{\aletter \bletter} \}$ and $\window^{\aletter \bletter}$ does not overlap with $\window^{\aletter \cletter}$, so no window merges occur. }{\linesref{AMPPforallcollidingwindows}{AMPPRemoveOldWindow}.}

  \statementjustif{\callsmall{\planin} is invoked to generate a repair in $\window^{\aletter \cletter}$. $\window^{\aletter \cletter}$ is added to the window set $W$.}{\linesref{AMPPPlanJointly}{AMPPAddWindow}. Shown in \figref{amppwindowrepair2}.}

  \statementjustif{No more collisions exist so \callsmall{\firstcollisionwindow} returns $\emptyset$ and the collision detection loop exits.}{\lineref{AMPPforallcollidingwindows}.}

  \statementjustif{$\window^{\aletter \bletter}$ does not allow for an unimpeded search from $\filterpath{\start}{\{ \aletter, \bletter \}}$ to $\filterpath{\goal}{\{ \aletter, \bletter \}}$, and $\window^{\aletter \cletter}$ does not allow for an unimpeded search from $\filterpath{\start}{\{ \aletter, \cletter \}}$ to $\filterpath{\goal}{\{ \aletter,  \cletter \}}$, so \callsmall{\shouldquit} returns false for both windows and $W$ remains unchanged.}{\linesref{AMPPShouldQuitLoop}{AMPPShouldQuit}.}

\end{tabular}

\input{ampp_example_figure.tex}

\noindent
\begin{tabular}{p{25.5em}|p{7.5em}}

  \statementjustif{$W$ is not empty so the global path is not returned as optimal, but it is reported as an intermediary solution along with its optimality bound.}{\linesref{AMPPReportDone}{AMPPReportIntermediary}.}

  \statementjustif{\callsmall{\recampp{}} is recursively invoked, with $W = \{\window^{\aletter \bletter}, \window^{\aletter \cletter} \}$ and the valid but potentially suboptimal plan.}{\lineref{AMPPPlanJointly}.}

  \statementjustif{$\window^{\aletter \bletter}$ is grown and replanned in, producing a larger $\window^{\aletter \bletter}$ and a repair. The larger $\window^{\aletter \bletter}$ replaces its predecessor in $W$, and it does not overlap with $\window^{\aletter \cletter}$ so they do not merge. }{\linesref{AMPPGrowAndReplan}{AMPPGrowAndReplanEnd}.}

  \statementjustif{$\window^{\aletter \cletter}$ is grown, and its successor overlaps with $\window^{\aletter \bletter}$, so $\window^{\aletter \cletter}$ is removed from $W$ such that $W = \{ \window^{\aletter \bletter}\}$, and $\window^{\aletter \cletter}$'s successor is to be merged with $\window^{\aletter \bletter}$. }{\linesref{AMPPCheckInterfere}{AMPPOverlapMerge}. Shown in \figref{amppexamplewindowsoverlap}.}

  \statementjustif{As $\window^{\aletter \bletter}$ and $\window^{\aletter \cletter}$ overlap, \callsmall{\mergecollidingwindows} is invoked. These windows are merged together to form $\window^{\aletter \bletter \cletter}$ and a repair is generated inside it.  $\window^{\aletter \bletter \cletter}$ is added to $W$, replacing $\window^{\aletter \bletter}$ and $\window^{\aletter \cletter}$ such that $W = \{ \window^{\aletter \bletter \cletter} \}$ .}{\linesref{AMPPforallcollidingwindows}{AMPPMergeCollideReturns}. Shown in \figref{amppexamplewindowmerge}.}

  \statementjustif{No collisions exist  and $\window^{\aletter \bletter \cletter}$ does not allow for an unimpeded search from $\filterpath{\start}{\{ \aletter, \bletter, \cletter \}}$ to $\filterpath{\goal}{\{ \aletter, \bletter, \cletter \}}$, so the updated global path is reported as an intermediary solution and \callsmall{\recampp} is recursively invoked.}{ \linesref{AMPPWhileExistsCollision}{AMPPReportDone}.}

  \statementjustif{\callsmall{\recampp} proceeds, growing $\window^{\aletter \bletter \cletter}$ and updating its repair and intermediary solutions. No collisions are introduced and $\window^{\aletter \bletter \cletter}$ does not allow for an unimpeded search from $\filterpath{\start}{\{ \aletter, \bletter, \cletter \}}$ to $\filterpath{\goal}{\{ \aletter, \bletter, \cletter \}}$ }{\linesref{AMPPGrowAndReplanStart}{AMPPRecurse}.}

  \statementjustif{\callsmall{\recampp} proceeds, growing $\window^{\aletter \bletter \cletter}$ and updating its repair and intermediary solutions, with no collisions introduced. The repair in $\window^{\aletter \bletter \cletter}$  allowed for an unimpeded search from $\filterpath{\start}{\{ \aletter, \bletter, \cletter \}}$ to $\filterpath{\goal}{\{ \aletter, \bletter, \cletter \}}$, therefore allowing \callsmall{\shouldquit} to return true. This removes $\window^{\aletter \bletter \cletter}$ from $W$, making $W$ empty and thus returns returns the global path as optimal. }{\linesref{AMPPGrowAndReplanStart}{AMPPReportDone}. Shown in \figref{amppexamplewindowmax}.}
  \hline
\end{tabular}

\subsubsection{Globally Suboptimal Repairs and Path Padding Example}\subsectionlabel{pathpaddingexample}

The example shown in \figref{ampppaddingexample} demonstrates how \ampp{} can produce a globally suboptimal path from an optimal repair within a window, and how higher quality repairs are padded to prevent breaking the entry state of windows further along the path using NWA*'s window definition. The associated figures first demonstrate an initial collision caused by \ampp{} planning individually (\figref{ampppaddingfigone}). \ampp{} then creates a repair window $\window^{\aletter \bletter}$ that is searched to find an repair, forcing agent $\bletter$ to step inside the slot in the wall to let $\aletter$ pass, thus adding two more moves to the global path cost. Due to $\window^{\aletter \bletter}$'s constraints, the repair was unable to consider instead sending $\aletter$ above the upper wall, towards its goal which would produce no increase in global path cost; as such, the repair generated is optimal within $\window^{\aletter \bletter}$ but produces to a suboptimal global path. This path also causes a collision later along the path (\figref{ampppaddingfigtwo}) that is then repaired with a second window $\window^{\bletter \cletter}$ and \ampp{} returns a valid solution (\figref{ampppaddingfigthree}). 

\input{ampp_padding.tex}

\noindent
Finally, the first window is grown, producing the repair of sending $\aletter$ above the upper wall and allowing $\bletter$ to travel without stepping into the slot; however, this improved repair would cause $\bletter$ to arrive at $\window^{\bletter \cletter}$ window too early as compared to its prior plan; in order to prevent this invalidation, the repair to $\bletter$ is padded with two waits to ensure that $\bletter$ leaves $\window^{\aletter \bletter}$ at the same time as it did previously in order to ensure it arrives at $\window^{\bletter \cletter}$ at the proper time. (\figref{ampppaddingfigfour}). By performing this padding, \ampp{} ensures all agents leave the window at the times they did previously, thus guaranteeing leaving agents will travel the same paths and enter later windows as they did previously, thereby avoiding the introduction of any new collisions or invalidation of later windows and thus quickly generating successive solutions. Ultimately, $\window^{\aletter \bletter }$ and $\window^{\bletter \cletter}$ will merge, absorbing the padded section of $\bletter$ and allowing for the optimal global path to be generated.

\input{ampp_arb_graph.tex}

\subsubsection{\ampp{} In Domains Without Regular Structure}

While the underlying planners for \ampp{} may exploit additional domain structure, \ampp{} itself exploits domain structure by using the window definition to carve the graph into small, self-contained collision repair problems. To do this effectively , the problem itself must be \emph{sparse}, i.e.\ amenable to this carving approach, \emph{and} \ampp{} must be provided with a window definition that effectively performs the carving process. As defined in \subsectionref{nwa}, NWA*'s window definition uses the regular structure of four-connected grids to compactly define such a window via a hyper-rectangle. In order to generalize to other regular grids, e.g.\ a hexagonal grid, this definition can be augmented to fit the grid's regular shape, e.g.\ a hyper-hexagon, and in order to generalize to an arbitrary graph with no known additional structure, this definition can be augmented to all states at most $k$ degrees of separation away from one or more center states. This fully general definition is shown in \figref{ampparbgraph}; while the graph has $L_2$ cost edges for ease of presentation, \ampp{} knows nothing beyond the graph's fundamental definition and is still able to operate. Alternative general window definitions include adding the state least expensive to reach from a center state and outside of the window, or the set of neighbors culled during the previous repair search by the existing window's constraints (this is provided for free by X*'s \emph{out-of-window set}, presented in \sectionref{xstar}).

\section{Expanding A*}\sectionlabel{xstar}

Expanding A* (X*) is an efficient \ampp{}-based planner. X* is nearly identical to NWA*~(\subsectionref{nwa}), differing only in implementing additional bookkeeping to allow re-use of prior repair search information when solving for a successive repair. Due to this re-use, X* is significantly more efficient than NWA* for successive plan generation. As we demonstrate empirically in \sectionref{results}, in sparse domains X* outperforms the state-of-the-art in time to first solution while remaining competitive with the state-of-the-art in time to optimal solution.

\subsection{X*'s Bookkeeping and Search Re-Use for Successive Plan Generation}\subsectionlabel{xstaroverview}

X*'s bookkeeping during the search for a repair in the window $\windowk$ allows for the resulting search tree to be transformed into a search tree in $\windowkpone$, saving computation during successive planning. The intuition behind X*'s bookkeeping and transformations is depicted in \figref{XStarSteps} as a \projectedillustration{}, i.e.\ a two-dimensional illustration depicting the higher-dimensional joint space of $\windowk$.

\newcommand{\figuretextsize}{\small}
\newcommand{\teasertextsize}{\scriptsize}
\newcommand{\stagefigscale}{0.675}
\newcommand{\outofwindowlistthickness}{0.15}
\definecolor{opensetcolor}{RGB}{137, 207, 240}
\definecolor{extendedopensetcolor}{RGB}{37, 107, 140}

\begin{figure}[H]
  \captionsetup[subfigure]{justification=centering}
  \centering
  \begin{subfigure}[t]{0.22\textwidth}
    \centering
    \begin{tikzpicture}[scale=0.4]
      \tikzstyle{every node}+=[inner sep=0pt]
        \draw[fill=opensetcolor, draw=none]
        [path picture]
        (0, 2.5)coordinate to (4, 3) coordinate to (4, 1.5) coordinate to (3, 0) coordinate to (1.75, 0) to cycle;        
      \draw [color=darkgreen] (0, 2.5) -- (1,2.2) -- (1.5, 1.5) -- (3,2.1) -- (4, 1.5);
      \draw [color=orange] (-1.5, 4) -- (-1, 3) -- (0, 2.5);
      \draw [color=orange] (4, 1.5) -- (5, 0.5) -- (5.5, -0.5);
      \draw [line width=\windowwidth] (0,0) rectangle (4,4);
      \draw [line width=\windowwidth] (-1, -1) rectangle (5,5); 
      \draw [draw=none, fill=red] (0, 2.5) circle (0.15cm); 
      \draw [draw=none, fill=green] (4, 1.5) circle (0.15cm); 
      \node at (0.75, 0.4) {\figuretextsize$\windowk$};
    \node at (0, -0.6) {\figuretextsize$\windowkpone$};
      \node at (0.45, 3) {\figuretextsize$\startk$};
      \node at (4.6, 1.7) {\figuretextsize$\goalk$};
      \draw [->,very thick] plot [smooth, tension=1] coordinates {(3, 5.7) (6, 6.2) (9,5.7)};
    \end{tikzpicture}
    \caption{\initialconfig{}}
    \figlabel{XStarStepsInitialConfig}
  \end{subfigure}%
  \hfill
\begin{subfigure}[t]{0.22\textwidth}
  \centering
  \begin{tikzpicture}[scale=0.4]
    \tikzstyle{every node}+=[inner sep=0pt]
    \draw[fill=opensetcolor, draw=none]
      [path picture]
      (0, 2.5) coordinate to (5, 3.125) coordinate to (3-0.666*0.91, -1*0.91) coordinate to cycle; 
      \draw [color=darkgreen] (0, 2.5) -- (1,2.2) -- (1.5, 1.5) -- (3,2.1) -- (4, 1.5);
      \draw [color=orange] (-1.5, 4) -- (-1, 3) -- (0, 2.5);
      \draw [color=orange] (4, 1.5) -- (5, 0.5) -- (5.5, -0.5);
    \draw[\windowstyle, line width=\windowwidth] (0,0) rectangle (4,4);
    \draw [line width=\windowwidth] (-1, -1) rectangle (5,5); 
    \draw [draw=none, fill=red] (0, 2.5) circle (0.15cm); 
    \draw [draw=none, fill=green] (4, 1.5) circle (0.15cm); 
    \node at (0.75, 0.4) {\figuretextsize$\windowk$};
  \node at (0, -0.6) {\figuretextsize$\windowkpone$};
    \node at (0.45, 3) {\figuretextsize$\startk$};
    \node at (4.6, 1.7) {\figuretextsize$\goalk$};
    \draw [->,very thick] plot [smooth, tension=1] coordinates {(3, 5.7) (6, 6.2) (9,5.7)};
  \end{tikzpicture}
  \caption{\stageone{:}\\{\footnotesize \stageonename{}}}
  \figlabel{XStarStepsStageOne}
\end{subfigure}%
\hfill
\begin{subfigure}[t]{0.22\textwidth}
  \centering
  \begin{tikzpicture}[scale=0.4]
    \tikzstyle{every node}+=[inner sep=0pt]
    \draw[fill=opensetcolor, draw=none]
      [path picture]
      (-1, 3) coordinate to (5, 3.125) coordinate to (3-0.666*0.91, -1*0.91) coordinate to cycle; 
      \draw [color=darkgreen] (-1, 3) -- (1,2.2) -- (1.5, 1.5) -- (3,2.1) -- (4, 1.5);
      \draw [color=orange] (-1.5, 4) -- (-1, 3);
      \draw [color=orange] (4, 1.5) -- (5, 0.5) -- (5.5, -0.5);
    \draw[\windowstyle, line width=\windowwidth] (0,0) rectangle (4,4);
    \draw [line width=\windowwidth] (-1, -1) rectangle (5,5); 
    \draw [draw=none, fill=red] (-1, 3) circle (0.15cm); 
    \draw [draw=none, fill=green] (4, 1.5) circle (0.15cm); 
    \node at (0.75, 0.4) {\figuretextsize$\windowk$};
  \node at (0, -0.6) {\figuretextsize$\windowkpone$};
    \node at (-1 + 0.95, 3.59) {\figuretextsize$\startkpone$};
    \node at (4.6, 1.7) {\figuretextsize$\goalk$};
    \draw [->,very thick] plot [smooth, tension=1] coordinates {(3, 5.7) (6, 6.2) (9,5.7)};
  \end{tikzpicture}
  \caption{\stagetwo{:}\\{\footnotesize \stagetwoname{}}}
  \figlabel{XStarStepsStageTwo}
\end{subfigure}%
\hfill
\begin{subfigure}[t]{0.22\textwidth}
  \centering
  \begin{tikzpicture}[scale=0.4]
    \tikzstyle{every node}+=[inner sep=0pt]
    \draw[fill=opensetcolor, draw=none]
      [path picture]
      (-1, 3) coordinate to (5, 3.125) coordinate to (5, 0.5) coordinate to (4, -1) coordinate to (2.333, -1) coordinate to cycle; 
      \draw [color=darkgreen] (-1, 3) -- (1,2.2) -- (1.5, 1.5) -- (3,2.1) -- (5, 0.5);
      \draw [color=orange] (-1.5, 4) -- (-1, 3);
      \draw [color=orange] (5, 0.5) -- (5.5, -0.5);
    \draw[\windowstyle, line width=\windowwidth] (0,0) rectangle (4,4);
    \draw [line width=\windowwidth] (-1, -1) rectangle (5,5); 
    \draw [draw=none, fill=red] (-1, 3) circle (0.15cm); 
    \draw [draw=none, fill=green] (5, 0.5) circle (0.15cm); 
    \node at (0.75, 0.4) {\figuretextsize$\windowk$};
  \node at (0, -0.6) {\figuretextsize$\windowkpone$};
    \node at (-1 + 0.95, 3.59) {\figuretextsize$\startkpone$};
    \node at (3.97, 0.57) {\figuretextsize$\goalkpone$};
  \end{tikzpicture}
  \caption{\stagethree{:}\\{\footnotesize \stagethreename{}}}
  \figlabel{XStarStepsStageThree}
\end{subfigure}%
\caption{\projectedillustration{s} of the three stage transformation employed by X* to enable search tree re-use. $\windowk$ and $\windowkpone$ represent the $k$th and $k+1$th windows, respectively. $\startk$ and $\startkpone$ represent the repair start for $\windowk$ and $\windowkpone$, respectively. $\goalk$ and $\goalkpone$ represent the repair goal for $\windowk$ and $\windowkpone$, respectively. \initialconfig{} (\figref{XStarStepsInitialConfig}) show the initial search tree. \stageone{} (\figref{XStarStepsStageOne}) grows the window without moving the start or goal. \stagetwo{} (\figref{XStarStepsStageTwo}) moves the start while keeping the same goal. \stagethree{} (\figref{XStarStepsStageThree}) moves the goal.}
\figlabel{XStarSteps}
\end{figure}%

\subsubsection{Search Re-Use: An A* Perspective}\subsectionlabel{astarperspective}

The three transformations depicted in \figref{XStarSteps} take an A*-style Search Tree from a repair search in $\windowk$ (\figref{XStarStepsInitialConfig}) that produced an optimal repair in $\windowk$ and transform it into a A*-style Search Tree for a repair search in $\windowkpone$ (\figref{XStarStepsStageThree}), producing an optimal repair in $\windowkpone$. 

\paragraph{\initialconfig{}}
The initial state, \emph{\initialconfig{}} (\figref{XStarStepsInitialConfig}), depicts a search tree from $\startk$ to $\goalk$ restricted inside $\windowk$; for now, we can imagine that this search tree was produced by standard A*. 

\paragraph{\stageone{}}
The first stage, \emph{\stageone{}: \stageonename{}} (\figref{XStarStepsStageOne}), depicts this search tree transformed to be as if the search took place from $\startk$ to $\goalk$ in the less restrictive $\windowkpone$. In order to go from an A* search tree in the smaller window $\windowk$ to a larger window $\windowkpone$, we need to expand all the states that would have been expanded in a search of $\windowkpone$ but are blocked by $\windowk$. These states, depicted in dark blue in \figref{bookkeepingexamplesneighbortracking}, must be reached via a state not in $\windowk$ whose direct predecessor is in $\windowk$; the set of these states is depicted in yellow in \figref{bookkeepingexamplesneighbortracking}. This motivates our \textbf{first bookkeeping addition: \emph{out of window set}}. For each state $\state \in \windowk$ that was expanded, we keep track of the neighbors of $\state$ that were discarded due to the restrictions of $\windowk$, i.e. $\neighbors{\state} \setminus \windowk$, placing them the \emph{out of window set}. This bookkeeping allows us to add these states to A*'s open set $\openset$, thereby initializing the search frontier in $\windowkpone$, depicted in yellow in \figref{bookkeepingexamplesneighbortracking}, Additionally, this bookkeeping provides a convenient way to track if the search was impeded when computing \callsmall{\shouldquit}; if the out of window set is empty after a repair search in $\windowk$, then the search in $\windowk$ was unimpeded.

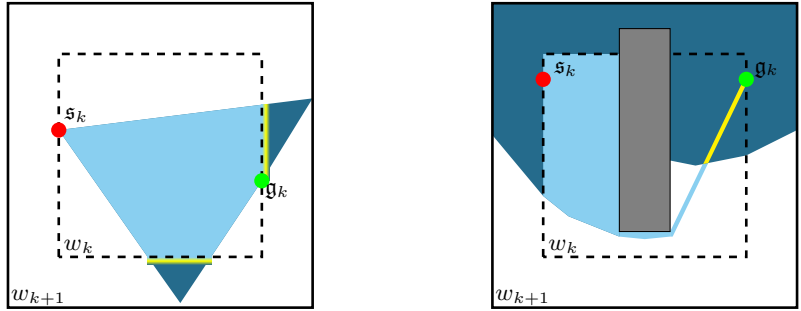
\begin{figure}
  \centering
  \begin{subfigure}[t]{0.47\textwidth}
    \centering
    \begin{tikzpicture}[scale=\stagefigscale]
      \tikzstyle{every node}+=[inner sep=0pt]
      \draw[fill=extendedopensetcolor, draw=none]
        [path picture]
        (0, 2.5)coordinate to (5, 3.125) coordinate to (4, 1.5) coordinate to (3-0.666*0.91, -1*0.91) coordinate to (1.75, 0) to cycle;
        \draw[fill=opensetcolor, draw=none]
        [path picture]
        (0, 2.5)coordinate to (4, 3) coordinate to (4, 1.5) coordinate to (3, 0) coordinate to (1.75, 0) to cycle;        
      \shade[shading=axis,bottom color=extendedopensetcolor,top color=opensetcolor,middle color=yellow,shading angle=0] (1.75,0) rectangle (3,-\outofwindowlistthickness);
      \begin{scope}[transform canvas={rotate=90}] 
          \shade[shading=axis,bottom color=extendedopensetcolor,top color=opensetcolor, middle color=yellow,shading angle=0] (3,-4) rectangle (1.5, -4-\outofwindowlistthickness);
      \end{scope}
      \draw[\windowstyle, line width=\windowwidth] (0,0) rectangle (4,4);
      \draw [line width=\windowwidth] (-1, -1) rectangle (5,5); 
      \draw [draw=none, fill=red] (0, 2.5) circle (0.15cm); 
      \draw [draw=none, fill=green] (4, 1.5) circle (0.15cm); 
      \node at (0.4, 0.2) {\figuretextsize$\windowk$};
    \node at (-0.4, 0.2-1) {\figuretextsize$\windowkpone$};
      \node at (0.35, 2.8) {\figuretextsize$\startk$};
      \node at (4.3, 1.3) {\figuretextsize$\goalk$};
    \end{tikzpicture}
    \caption{First bookkeeping addition example. Yellow region shows states stored in out of window set during the search of $\windowk$, where the light blue region was expanded. These yellow states form the frontier for the expansion of states when the window is grown to $\windowkpone$ and the search encompasses the dark blue area.}
    \figlabel{bookkeepingexamplesneighbortracking}
  \end{subfigure}%
  \hfill
\begin{subfigure}[t]{0.47\textwidth}
  \centering
  \begin{tikzpicture}[scale=\stagefigscale]
    \tikzstyle{every node}+=[inner sep=0pt]
    \draw[fill=opensetcolor, draw=none]
    [path picture]
    (2.5, 0.4) to (2.55, 0.4) to (4, 3.4) to (4, 3.6) to (2.5, 0.5) to cycle;
    \node at (1.95, 4.8) {\tiny{} Hello World! Thanks for reading this!};
    \node at (1.95, 4.55) {\tiny{} I worked very hard on this paper};
    \node at (1.95, 4.3) {\tiny{} and I am very proud of the results!};
    \node at (1.95, 4.05) {\tiny{} One day I will show this to my kids!};
    \draw[fill=extendedopensetcolor, draw=none]
    [path picture]
    (-1, 3.5) to (-1, 5) to (5, 5) to (5, 2.5) to (4, 2) to (3, 1.8) to (2.5, 1.9) to (1.5, 0.4) to (0.5, 0.8) to (0, 1.2) to (-1, 2.4) to cycle;
    \draw[fill=opensetcolor, draw=none]
    [path picture]
    (0, 3.5) to (0, 4) to (1.5, 4) to (1.5, 0.4) to (0.5, 0.8) to (0, 1.2) to cycle;        
    \tikzstyle{every node}+=[inner sep=0pt]
    \draw[fill=opensetcolor, draw=none]
    [path picture]
    (1.5, 0.4) to (2, 0.35) to (2.5, 0.4) to (2.5, 0.5) to (1.5, 0.5) to cycle;        
    \tikzstyle{every node}+=[inner sep=0pt]
    \tikzstyle{every node}+=[inner sep=0pt]
    \draw[fill=yellow, draw=none]
    [path picture]
    (3.15, 1.83) to (3.25, 1.85) to (4, 3.4) to (4, 3.6) to cycle;

    %
    \draw [\windowstyle, line width=\windowwidth] (0,0) rectangle (4,4); 
    \draw [line width=\windowwidth](-1, -1) rectangle (5,5); 
    \draw [fill=gray] (1.5, 4.5) rectangle (2.5,0.5); 
    \draw [draw=none, fill=red] (0, 3.5) circle (0.15cm); 
    \draw [draw=none, fill=green] (4, 3.5) circle (0.15cm); 
    \node at (0.4, 0.2) {\figuretextsize$\windowk$};
    \node at (-0.4, 0.2-1) {\figuretextsize$\windowkpone$};
    \node at (0.45, 3.7) {\figuretextsize$\startk$};
    \node at (4.4, 3.7) {\figuretextsize$\goalk$};
  \end{tikzpicture}
  \caption{Second bookkeeping addition example. Gray object is a joint space obstacle. Light blue region indicates area expanded during initial search of $\windowk$ to $\goalk$. Dark blue region indicates area expanded during expansion of states in $\windowkpone$ based on the \fvalue{} of $\goal$'s expansion in $\windowk$. Yellow region indicates states  re-expanded with a lower \gvalue{}.}
  \figlabel{bookkeepingexamplesreexpansion}
\end{subfigure}%
\caption{\projectedillustration{s} of motivating examples for the two bookkeeping additions.}
\figlabel{bookkeepingexamples}
\end{figure}%

When the window is grown, we also need to consider the possibility of new, shorter paths to already expanded states. An example of this is shown in \figref{bookkeepingexamplesreexpansion}, where the gray obstacle forces a search constrained by $\windowk$ to travel below it to reach $\goalk$, but a search in $\windowkpone$ allows for travel above the gray obstacle to not only reach $\goalk$ more quickly, but also more quickly reach the other states depicted in yellow.
As such, we must allow for states which were expanded in the search of $\windowk$ to be re-expanded in the search of $\windowkpone$ if the search in $\windowkpone$ assigns these states a lower \gvalue{}. This motivates our \textbf{second bookkeeping addition: \emph{closed value}}. In order to facilitate this re-expansion, during the initial A* search we also track the \gvalue{} at which a state is placed into the closed set $\closedset$, called the state's \emph{closed value}. 

It is important to note that all states in $\closedset$ at the end of the search of $\windowk$ cannot be reached with a lower cost than their closed value via any path that stays entirely within $\windowk$; as the search in $\windowk$ is optimal, any lower cost path to any state in $\closedset$ must leave $\windowk$, travel through a portion of $\windowkpone$, and re-enter $\windowk$, just as the path above the gray obstacle did in \figref{bookkeepingexamplesreexpansion}. Thus, the addition of the states to $\openset$ from our out of window set (first bookkeeping addition) ensures that all of such paths are able to be considered as long as states are able to be re-expanded if their closed \gvalue{}, recorded by our closed value (second bookkeeping addition), is higher than their \gvalue{} as they sit in $\openset$. With this modification, we can run A* until the minimal \fvalue{} in $\openset$ is greater than the \fvalue{} of $\goalk$. This will update all of the states in $\closedset$ and $\openset$ to have the optimal \gvalue{} for a search in $\windowkpone$ and thus produce the search tree shown in \stageone{}.

\begin{figure}
  \centering
    \begin{tikzpicture}[scale=\stagefigscale]
      \tikzstyle{every node}+=[inner sep=0pt]
      \draw[fill=opensetcolor, draw=none]
        [path picture]
        (2, 1.5)coordinate to (6, 2) coordinate to (6, 0) coordinate to (3.5, 0) to cycle;
      
      \draw[\windowstyle, line width=\windowwidth] (2,0) -- (2,4); 
      \draw[\windowstyle, line width=\windowwidth] (2,4) -- (6,4); 

      \draw[orange] (0, 2.5) -- (2, 1.5);
      \draw[darkgreen] (5, 1.7) -- (2, 1.5);
      \draw[blue] (5, 1.7) -- (0, 2.5);

      \draw [line width=\windowwidth] (0,0) -- (0,6); 
      \draw [line width=\windowwidth] (0,6) -- (6,6); 
      
      \draw [draw=none, fill=red] (0, 2.5) circle (0.15cm); 
      \draw [draw=none, fill=red] (2, 1.5) circle (0.15cm); 
      \draw [draw=none, fill=yellow] (5, 1.70) circle (0.15cm); 
      \node at (5.7, 3.75) {\figuretextsize$\windowk$};
      \node at (5.4, 5.75) {\figuretextsize$\windowkpone$};
      \node at (2.3, 1.8) {\figuretextsize$\startk$};
      \node at (0.5, 2.8) {\figuretextsize$\startkpone$};
      \node at (5.25, 1.5) {\figuretextsize$\state'$};
    \end{tikzpicture}
    \caption{\projectedillustration{} of moving the search tree start from $\startk$ to $\startkpone$ along a section of the joint space between the two starts depicted in orange. The yellow point $\state'$ can be reached from $\startkpone$ by traveling from $\startkpone$ to $\startk$ along the joint space path and then, using the information from the search tree shown in light blue, travel from $\startk$ to $\state'$ as depicted in green. It may not always the case that this is a minimal cost path from $\startkpone$ to $\state'$, as there may be a shorter path from $\startkpone$ to $\state'$ without traveling through $\startk$ such as the path depicted in blue, thus motivating a need to re-expansion of some states in the search tree.\vspace{-1.1em}}
    \figlabel{extendpathexample}
\end{figure}%

The second stage, \emph{\stagetwo{}: \stagetwoname{}} (\figref{XStarStepsStageTwo}), depicts the search tree transformed from a start $\startk$ as seen in \stageone{} to a start $\startkpone$. In order to move the start backwards, we need to embed the search tree rooted at $\startk$ into the search tree rooted at $\startkpone$. As illustrated in \figref{extendpathexample}, to reach any state in the existing tree from $\startkpone$, e.g. $\state'$ (depicted in yellow), the cost of the minimal path (depicted in blue) is upperbounded by the cost to travel from $\startkpone$ to $\start$ (depicted in orange) plus the cost to travel from $\startk$ to $\state'$ (depicted in green). This holds because the orange path from $\startkpone$ to $\startk$ is extracted from the global path $\pth$, which is provably collision-free in this region (see \appendixref{xstar}, \theoremref{Stage2OptCollisionFree} for proof), and thus serves a valid upperbound, and the green path from $\startk$ to $\state'$ is provided by the \gvalue{}s of the existing search tree and thus is the optimal cost from $\startk$ to $\state'$. Thus, if we increase every state's \gvalue{} and closed value (second bookkeeping addition) by the cost of the path from $\startkpone$ to $\startk$, and we expand each state along the path from $\startkpone$ to $\startk$, we can run A* until the minimal \fvalue{} in $\openset$ is greater than the \fvalue{} of $\goalk$, leveraging the second bookkeeping addition to re-expand states in the $\startk$ rooted tree as needed, as done in \stageone{}.

The third stage, \emph{\stagethree{}: \stagethreename{}} (\figref{XStarStepsStageThree}), depicts the search tree rooted at $\startkpone$ transformed from a goal $\goalk$ as seen in \stagetwo{} to a goal $\goalkpone$. The states in $\openset$ simply need to have their \fvalue{s} updated with new \hvalue{s} to $\goalkpone$ and then A* can be run as normal until $\goalkpone$ is expanded. 

\subsubsection{Bookkeeping Formalization}\subsectionlabel{bookkeepingformalisms}

In order to be able to reason about a state's \fvalue{}, \gvalue{}, and \hvalue{} under different starts and goals, we augment the $f$, $g$, and $h$ function with start and goal parameters. For example, given a state $\state$, start $\startk$, and goal $\goalk$, $\state$'s \fvalue{}, \gvalue{}, and \hvalue{} are $\f{\state}{\startk}{\goalk}$, $\g{\state}{\startk}$, and $\h{\state}{\goalk}$, respectively. Like standard A*, if any \gvalue{} entry has not been set, it returns $\infty$.

\subsectionref{astarperspective} discusses two bookkeeping additions to standard A* search trees that facilitate the search re-use depicted in \figref{XStarSteps}. The first bookkeeping addition, called an \emph{out of window} set $\outofwindowset$, maintains a set of all states that are neighbors of expanded states in $\windowk$ and themselves are not in $\windowk$. In \stageone{}, when $\windowk$ is grown to $\windowkpone$, the states $\{\state \mid \state \in \outofwindowset \cap \windowkpone \}$ are added to $\openset$ and removed from $\outofwindowset$. The second bookkeeping addition, called a state $\state$'s \emph{closed cost}, is recorded in $\cl{\state}{\start} \gets \g{\state}{\start}$ when $\state$ is placed in $\closedset$; this is similar to the bookkeeping done when running A* with an inconsistent heuristic~\cite{russellandnorvig}. This table is checked during state expansions in \stageone{} and \stagetwo{}'s transformations in order to allow the re-expansion of states which have shorter paths. Like \gvalue{s}, if an entry in $\clsymb$ has not been set, it returns $\infty$.

\subsection{\ampp{} Subroutine Implementations}\subsectionlabel{XStarImplementations}

Three of X*'s five key implementations are identical to NWA* (\subsectionref{nwa}); however,
the other two make use of the guarantees provided by \ampp{} regarding the
ordering of \callsmall{\planin}{} and \callsmall{\growandreplan}{} calls on successor
windows to improve efficiency. Additionally, for these re-use techniques to work, we assume the heuristic is consistent, i.e.\ the triangle inequality holds.

\amppcomponent{\Call{\planin}{$\window, \pth$}:} \sectionlabel{XStarPlanIn} This subroutine is implemented almost identically to NWA*'s \callsmall{\planin} in \subsectionref{nwa}, but with the implementation of the two bookkeeping additions from \subsectionref{bookkeepingformalisms}. \callsmall{A*WithBookkeeping} in \algoref{XStarGrowAndReplan} is A* modified with these bookkeeping additions.


  \amppcomponent{\Call{\growandreplan}{$\windowk, \pth$}:} As defined in \sectionref{\ampp}, \callsmall{\growandreplan}{} will only be invoked on a window in which \callsmall{\growandreplan}{} or \callsmall{\planin}{} were previously invoked. As such, this subroutine leverages the \xstarsearchtree{} produced by the previous search of $\windowk$ to aid the current search of $\windowkpone$ via the transformation shown in \figref{XStarSteps} and discussed in \subsectionref{xstaroverview}.
The algorithm and its supporting procedures are presented in \algoref{XStarGrowAndReplan}. 

\begin{algorithm}
  \caption{\growandreplan{}}\algolabel{XStarGrowAndReplan}
  \begin{algorithmic}[1]

    \Function{\growandreplan}{$\windowk, \pth$}    
    \State \Call{Stage1}{}\linelabel{XStarGrowAndReplanStage1} \Comment{Produces \stageone{} in \figref{XStarSteps}.}
    \State \Call{Stage2}{}\linelabel{XStarGrowAndReplanStage2}\Comment{Produces \stagetwo{} in \figref{XStarSteps}.}
    \State $\pth' \gets$ \Call{Stage3}{}\linelabel{XStarGrowAndReplanStage3} \Comment{Produces \stagethree{} in \figref{XStarSteps}.}
    \State Replace section of $\filterpath{\pth}{\agentset}$ from $\windowkpone$'s $\start$ to $\goal$ with $\pth'$ \linelabel{XStarGrowAndReplanUpdateExistingPath}
    \State \Return $ \left( \windowkpone, \pth  \right)$ \linelabel{XStarGrowAndReplanReturn}
    \EndFunction

    \Procedure{ExpandState}{$\state, \start$}
    \State $\closedset \gets \closedset \cup \{s\}$
    \State $\cl{s}{\start} \gets \g{s}{\start}$
    \State $\openset \gets \openset \cup \{n \mid n \in
    \neighbors{s}: n \in \window \}$
    \State $\outofwindowset \gets \outofwindowset \cup \{n \mid n \in
    \neighbors{s}: n \not\in \window \}$ 
    \ForAll{$n \in \neighbors{s}$}
    $\g{n}{\start} \gets \min(\g{n}{\start}, \g{s}{\start} + \cost{s}{n})$
    \EndFor
    \EndProcedure

    \Procedure{A*SearchUntil}{$\openset, \closedset, \outofwindowset, \window, \fmax{}$}
    \While{ $\f{\topopenset{\start}{\goal}}{\start}{\goal} \leq \fmax{}$ }\linelabel{HaltFMax}
    \State $s \gets \topopenset{\start}{\goal}$\linelabel{AStarSearchUntilGetTop}
    \State $\openset \gets \openset \setminus \{s\}$ \linelabel{AStarSearchUntilRemoveFromO}
    \If{$s \in \closedset \land \cl{s}{\start} \leq \g{s}{\start}$}\linelabel{ReexpandState}
    \Continue
    \EndIf
    \State \Call{ExpandState}{$\state, \start$}
    \EndWhile
    \EndProcedure

    \Procedure{A*WithBookkeeping}{$\openset, \closedset, \outofwindowset, \window, \start, \goal$}
    \While{ $\openset \neq \emptyset$ }\linelabel{Stage3AStarStart}
    \State $s \gets \topopenset{\start}{\goal}$
    \If{$s = \goal$}
    \Return \Call{UnwindPath}{$\closedset, \goal, \start$}
    \EndIf
    \State $\openset \gets \openset \setminus \{s\}$
    \If{$s \in \closedset$}
    \Continue  \linelabel{Stage3ClosedSet} \EndIf
    \State \Call{ExpandState}{$\state, \start$}
    \EndWhile
    \State \Return $\nopath{}$
    \EndProcedure

    \Procedure{Stage1}{}
    \State $\openset \gets \openset \cup \{s \mid s \in \outofwindowset: s
    \in \windowkpone\}$ \linelabel{Stage1AddOpenList}
    \State $\outofwindowset \gets \{s \mid s \in \outofwindowset: s
    \not\in \windowkpone\}$ \linelabel{Stage1FilterOutOfWindowList}
    \State $\Call{A*SearchUntil}{\openset, \closedset, \outofwindowset, \windowkpone, \f{\goalk}{\startk}{\goalk}}$\linelabel{Stage1AStarSearchUntil}
    \EndProcedure

    \Procedure{Stage2}{}
    \State $\pth' \gets$ path between $\startkpone$ and $\startk$ extracted from $\pth$\linelabel{Stage2ExtractPath}
     \ForAll{$s \in \openset \cup \closedset$}
    $\g{s}{\startkpone} \gets \g{s}{\startk} + \pthcost{\btwstartspath} $ \linelabel{Stage2IncreaseGValue}
     \EndFor
    \ForAll{$s \in \closedset$}
    $\cl{s}{\startkpone} \gets \cl{s}{\startk} + \pthcost{\btwstartspath}$\linelabel{Stage2IncreaseClosedValue}
    \EndFor
    \ForAll{$\state \in \pth'$} 
    \Call{ExpandState}{$\state, \startkpone$}\linelabel{Stage2Expand}
    \EndFor
    \State $\Call{A*SearchUntil}{\openset, \closedset, \outofwindowset, \windowkpone, \f{\goalk}{\startk}{\goalk} +
      \pthcost{\btwstartspath}}$\linelabel{Stage2AStarSearchUntil}
    \EndProcedure

    \Procedure{Stage3}{}
    \State Reorder $\openset$ using \fvalue{} from $\startkpone$ to $\goalkpone$. \Comment{\hvalue{s} changed}
    \If {$\goalkpone \in \closedset$}\linelabel{Stage3Reorder} 
    \Return \Call{UnwindPath}{$\closedset, \goalkpone, \startkpone$}\linelabel{Stage3UnwindEarly}
    \EndIf
    \State \Return \Call{A*WithBookkeeping}{$\openset, \closedset, \outofwindowset, \windowkpone, \startkpone, \goalkpone$}\linelabel{Stage3AStarBookkkeeping}
    \EndProcedure

      \end{algorithmic}
\end{algorithm}

\newcommand{\subroutinespace}{\vspace{-0.5em}}

\subroutinespace
\paragraph{\growandreplan} This algorithm performs setup and wraps three subroutines corresponding to the three stages shown in \figref{XStarSteps}. Importantly, on \lineref{XStarGrowAndReplanUpdateExistingPath}, X* replaces a section of the existing path $\pth$ with its repair $\pth'$. Due to the fact that we are growing an existing repair, $\pth$ is already a valid global path which we are improving. As such, we must ensure that if $\pth'$ is shorter than the existing region in $\pth$, $\pth'$ is padded so that all agents leave the state $\goalkpone$ at the same time they did in $\pth$; this is critical to ensuring any window repairs further along $\pth$ continue to have start states that are reachable from $\pth$.

\subroutinespace
\paragraph{\astarsearchuntil}

As discussed in \subsectionref{astarperspective}, \stageone{} and \stagetwo{} need to expand all states with less than or equal to a given \fvalue{} in order to ensure that states have the minimal cost \gvalue{} for the given window. \callsmall{A* Search Until}{} is a helper function which provides this functionality for a given \fvalue{}, $\fmax{}$, by running a modified A* search which only terminates when the minimal \fvalue{} of any state in $\openset$ is greater than $\fmax{}$. Note that the expansion skip condition for a state expansion (\lineref{ReexpandState}) also considers the closed value of the state, allowing for \callsmall{A* Search Until} to re-expand a state if its \gvalue{} is lower than its closed value.

\subroutinespace
\paragraph{A* With Bookkeeping} This procedure runs standard A* from the given start $\start$ to the given goal $\goal$ in the given window $\window$ using the given open set $\openset$, closed set $\closedset$, and out of window set $\outofwindowset$. Note that, unlike \callsmall{\astarsearchuntil}, the expansion skip condition for \callsmall{A*WithBookkeeping} is a standard A*-style $\closedset$ membership check (\lineref{Stage3ClosedSet}).

\subroutinespace
\paragraph{\stageone{}} This procedure converts the tree shown in \initialconfig{} (\figref{XStarStepsInitialConfig}) into \stageone{} (\figref{XStarStepsStageOne}). It does this by initializing $\openset$ with the frontier of the search for states in $\windowkpone$ but not in $\windowk$ and then leverages \callsmall{\astarsearchuntil} to expand or re-expand states with \fvalue{s} less than the \fvalue{} of $\goalk$, as these states would have been expanded during a direct search of $\windowkpone$.

\subroutinespace
\paragraph{\stagetwo{}}{} This procedure converts the tree shown in \stageone{} (\figref{XStarStepsStageOne}) into \stagetwo{} (\figref{XStarStepsStageTwo}). As discussed in \subsectionref{astarperspective}, it does this by extracting the relevant section of $\pth$ from $\startkpone$ to $\startk$(\lineref{Stage2ExtractPath}). The path cost is used to increase the \gvalue{} of each state in $\openset$ and $\closedset$ (\lineref{Stage2IncreaseGValue}), as well as the closed value of each state in $\closedset$ (\lineref{Stage2IncreaseClosedValue}). Then, all states along this path are expanded (\lineref{Stage2Expand}). Note that as each state's \gvalue{} in $\openset$ is increased by a fixed amount, no reordering of $\openset$ is required even if backed by an ordered data structure (e.g.\ a heap).

\subroutinespace
\paragraph{\stagethree{}}{} This procedure converts the tree shown in \stagetwo{} (\figref{XStarStepsStageTwo}) into \stagethree{} (\figref{XStarStepsStageThree}). As the goal moves from $\goalk$ to $\goalkpone$, the heuristic evaluation for each state in $\openset$ will change by differing amounts for various states and thus, if $\openset$ is backed by a structure such as a heap, it will require reordering (\lineref{Stage3Reorder}). The rest of \callsmall{\stagethree{}} is standard A* with bookkeeping additions  (\lineref{Stage3AStarBookkkeeping}).

\section{Empirical Results}\sectionlabel{results}

\newcommand{\examplegridcolor}{red}
\begin{figure}
  \compresscaptions
\centering
  \begin{subfigure}[b]{0.47\textwidth}
  \centering
  \input{example_grid.tex}
  \caption{5\% occupied $100 \times 100$ random grid.}
  \figlabel{example10percentrandom}
\end{subfigure}
\hfill
\begin{subfigure}[b]{0.47\textwidth}
  \centering
  \includegraphics[width=2in]{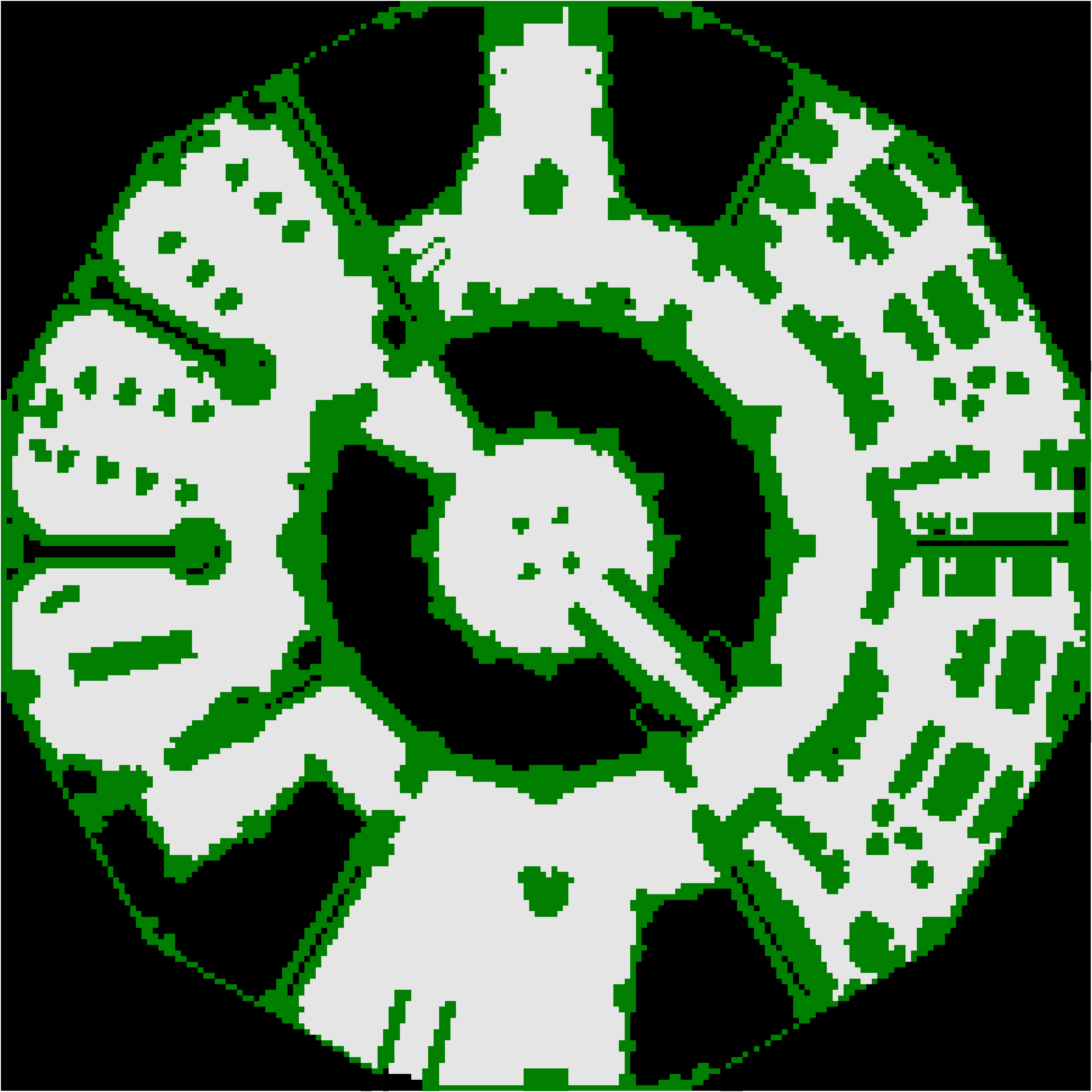}
  \caption{\texttt{lak303d} benchmark domain.}
  \figlabel{examplebenchmarkdomain}
\end{subfigure}
\caption{Examples of the domains in which experiments all experiments were performed.}
\figlabel{firstsolution1percent}
\end{figure}

In this section, we evaluate X* using randomly generated four-connected grids (e.g.\ \figref{example10percentrandom}) and several standard benchmark domains\footnote{\texttt{den520d},
\texttt{brc202d},
\texttt{lak303d},
\texttt{ht\_mansion\_n},
\texttt{ost003d},
and
\texttt{w\_woundedcoast} domains were used. Benchmarks available at
\url{https://movingai.com/benchmarks/mapf/index.html}. } (e.g.\ \figref{examplebenchmarkdomain}). All experiments treat the domains as uniform cost four-connected grids with randomly selected starts and goals. Unless stated otherwise, X* is configured with an initial window $L_{\infty} = 2$ and window expansions grow the window by a single step. All boxplot whiskers are at most the length of the interquartile range, with the lower whisker fit to the lowest datapoint above this value and the upper whisker fit to the highest datapoint below this value.  We use these domains in multiple experiments to evaluate:


\begin{enumerate}
  \item How X* compares to state-of-the-art \mapf{} planners in time to generate a valid path in sparse domains (\subsectionref{timetovalidsolution}).
  \item How X* compares to state-of-the-art \mapf{} planners in time to generate an optimal path in sparse domains (\subsectionref{timetooptimalsolution}).
  \item How X* compares to NWA* in valid path generation performance and optimal path generation performance (\subsectionref{XStarVsBaselines}).
  \item Which components of X* dominate its runtime (\subsectionref{XStarDominantComponenets}). 
  \item The effect of domain characteristics on the performance of X* (\subsectionref{XStarRuntimeVsSparsity}).
  \item Suboptimality bounds of X*'s first and intermediary paths (\subsectionref{suboptimalitybounds})
  \item The effect of parameters on the performance of X* (\subsectionref{XStarWindowSelection}).
\end{enumerate}



 All planners were implemented in C++. X* and NWA* were implemented by the
authors of this paper\footnote{Source code available at
  \url{https://github.com/kylevedder/libMultiRobotPlanning}}, AFS was
implemented by its original authors, CBS was implemented by
a third party\footnote{Source code available at
  \url{https://github.com/whoenig/libMultiRobotPlanning/}}, M* was
implemented by its original authors\footnote{Source code available at
  \url{https://github.com/gswagner/mstar_public/}}, (Operator
Decomposition version used), and PR was implemented by
a third party and modified by the authors of this paper\footnote{Source code available at
\url{https://github.com/kylevedder/Push-and-Rotate}}.
All runtime measurements were performed on a dedicated computer with an Intel i7 CPU (TurboBoost disabled) and access to 60GB of DDR4 RAM. Any trial that exceeded the memory limit was recorded as a timeout.

\subsection{Comparison for time to Valid Path}\subsectionlabel{timetovalidsolution}

In order to evaluate the performance of X* compared to state-of-the-art anytime or optimal \mapf{} solvers for time to valid path generation in sparse domains, we run X*, AFS, CBS, and M* with varied numbers of agents on randomly generated $100\times 100$ grids with 1\%, 5\%, and 10\% of the states blocked (\figref{validsolutiongrid}) and on standard benchmark domains for fixed number of agents (\tableref{CommonBenchmarks}, first rows).

\figref{validsolutiongrid} demonstrates that in random domains, X* outperforms the state-of-the-art \mapf{} planners in time to a valid path. X*'s improved performance is most distinct in domains with 1\% of states blocked, as these domains are especially sparse and thus amenable to X*'s approach; as the density of obstacles increases and thus domain sparsity decreases, the gap between X*'s performance and the state-of-the-art \mapf{} planners shrinks but is still pronounced. Compared to AFS and M*, X* on average produces a path at least an order of magnitude faster; while some of this performance difference may be the result of differing implementation quality, much of it can be attributed to the overhead of requiring a full joint search for AFS or individual space policy computations for M*. Compared to CBS, X* performs just as well for small numbers of agents, producing paths for 10 agents in under 10 milliseconds, but as the number of agents increases, performance diverges in favor of X*.

\tableref{CommonBenchmarks} reaffirms that X* is significantly faster than AFS and M* for time to valid path generation with over a two order of magnitude faster time, while CBS and X* are highly competitive; this result is a reflection of the high degree of sparsity in these domains and the initial overhead of AFS and M*.

Like all other planners in \figref{validsolutiongrid} and \tableref{CommonBenchmarks}, X* fails to generate a path in a reasonable amount of time for particularly challenging problems. As discussed in \subsectionref{XStarDominantComponenets}, this is caused by high dimensional searches resulting from repairs in windows with a large number of agents. 

In order to evaluate the performance of X* compared to suboptimal \mapf{} solvers for time to valid path generation in sparse domains, we run X* and PR with varied numbers of agents on randomly generated $100\times 100$ grids with 1\%, 5\%, and 10\% of the states blocked (\figref{prsolutions}). These results demonstrate that in random domains for small numbers of agents, X* outperforms PR in time to first path; while some of the performance difference can be attributed to implementation quality, much of it can be attributed to the fact that X* exploits the sparsity present in these test domains while PR does not. PR provides much more consistent runtimes in its valid path generation, solving all scenarios for all agent counts in under $\frac{1}{5}$th of a second and scaling quasi-linearly with increasing agent counts; however, PR provides significantly lower path quality than X*. PR's median path suboptimality factor, computed against an optimal path generated post-hoc, was (\texttt{2.0020}, \texttt{2.0673}, \texttt{2.1372}) across all runs for 1\%, 5\%, and 10\% obstacle density, respectively. X*'s online suboptimality factor, an exact or overestimate of the true suboptimality factor, was (\texttt{1.0029}, \texttt{1.0029}, \texttt{1.0029}) across all runs for 1\%, 5\%, and 10\% obstacle density, respectively (a full analysis of X*'s path quality bounds is presented in \subsectionref{suboptimalitybounds}). This experiment demonstrates X*'s advantage for time to valid path generation for small numbers of agents or when path quality is important.


\begin{figure}[htbp!]
  \compresscaptions
\centering
\begin{subfigure}[b]{0.3\textwidth}
  \centering
  \includegraphics[]{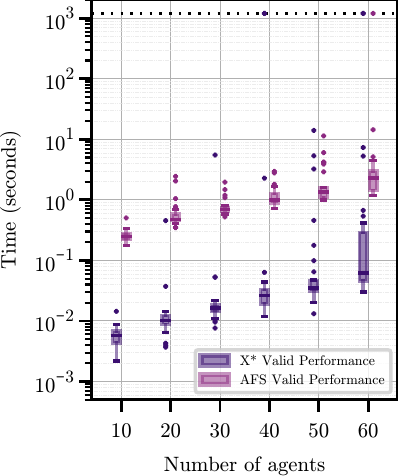}
  \caption{X* vs AFS 1\%}
\end{subfigure}
\hfill
  \begin{subfigure}[b]{0.3\textwidth}
  \centering
  \includegraphics[]{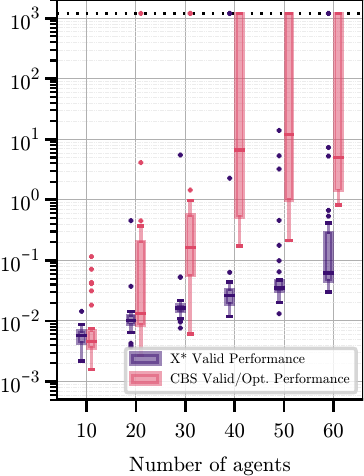}
  \caption{X* vs CBS 1\%}
\end{subfigure}
\hfill
\begin{subfigure}[b]{0.3\textwidth}
  \centering
  \includegraphics[]{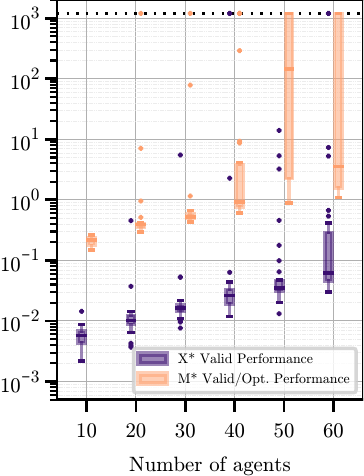}
  \caption{X* vs M* 1\%}
\end{subfigure}

\begin{subfigure}[b]{0.3\textwidth}
  \centering
  \includegraphics[]{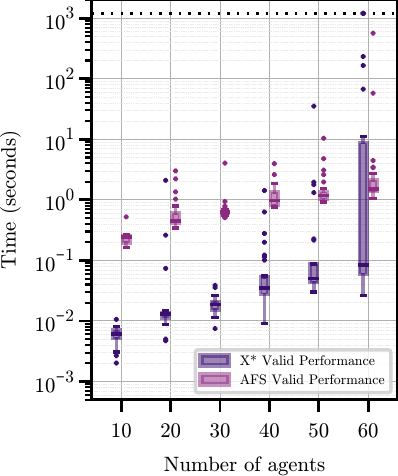}
  \caption{X* vs AFS 5\%}
\end{subfigure}
\hfill
  \begin{subfigure}[b]{0.3\textwidth}
  \centering
  \includegraphics[]{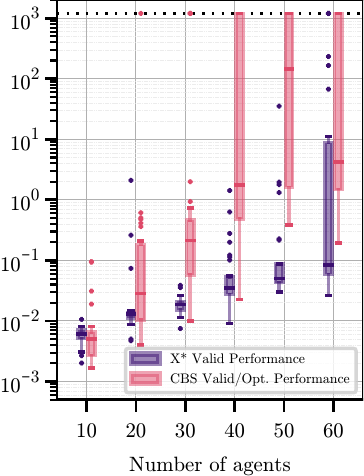}
  \caption{X* vs CBS 5\%}
\end{subfigure}
\hfill
\begin{subfigure}[b]{0.3\textwidth}
  \centering
  \includegraphics[]{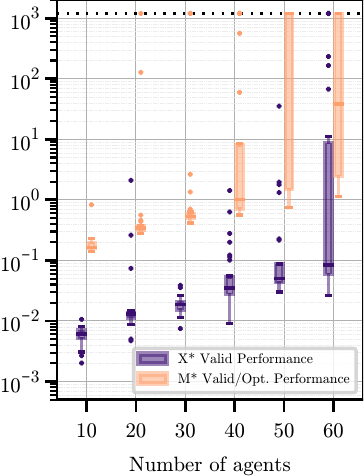}
  \caption{X* vs M* 5\%}
\end{subfigure}

\begin{subfigure}[b]{0.3\textwidth}
  \centering
  \includegraphics[]{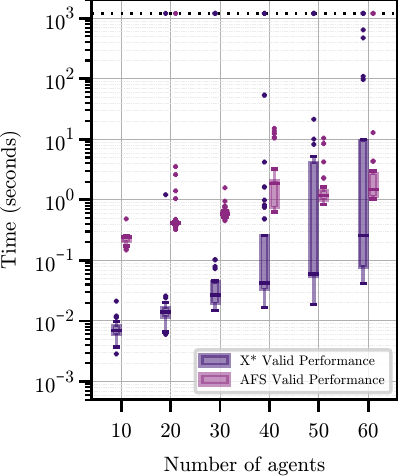}
  \caption{X* vs AFS 10\%}
\end{subfigure}
\hfill
  \begin{subfigure}[b]{0.3\textwidth}
  \centering
  \includegraphics[]{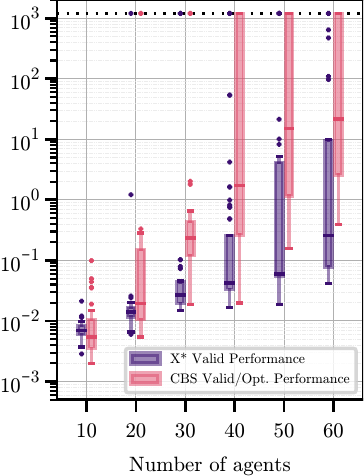}
  \caption{X* vs CBS 10\%}
\end{subfigure}
\hfill
\begin{subfigure}[b]{0.3\textwidth}
  \centering
  \includegraphics[]{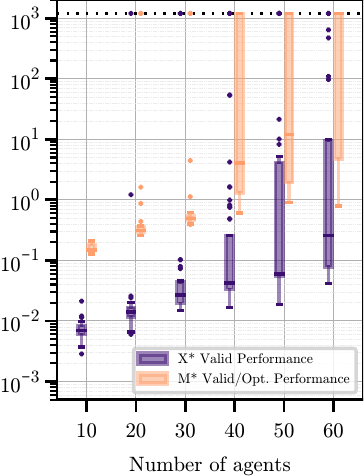}
  \caption{X* vs M* 10\%}
\end{subfigure}
\caption{Box plots of time to valid path
  for X*, AFS, CBS, and M* on a log scale. For each agents count, 30 trials are run, each with a 20 minute timeout, with
  each trial run on a randomly generated $100 \times 100$
  four-connected grid. The percentage of states blocked is listed in each caption.}
\figlabel{validsolutiongrid}
\end{figure}



\begin{figure}[htbp!]
  \compresscaptions
\centering
\begin{subfigure}[b]{0.33\textwidth}
  \centering
  \includegraphics[]{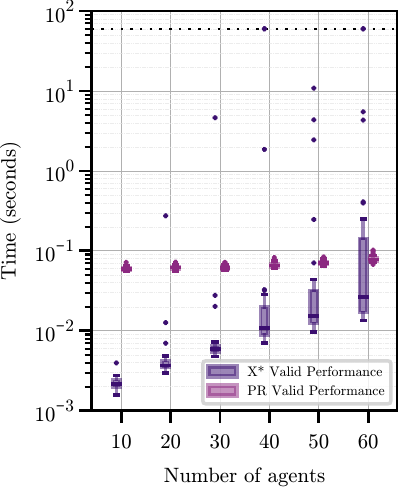}
  \caption{X* vs PR 1\%}
\end{subfigure}
\hfill
  \begin{subfigure}[b]{0.3\textwidth}
  \centering
  \includegraphics[]{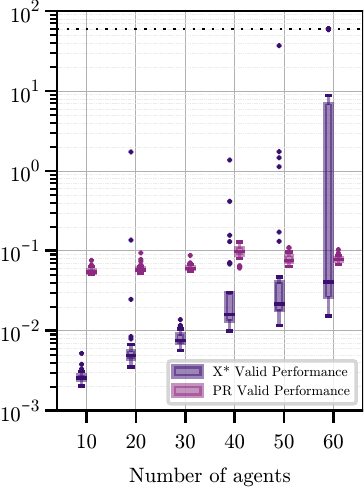}
  \caption{X* vs PR 5\%}
\end{subfigure}
\hfill
\begin{subfigure}[b]{0.3\textwidth}
  \centering
  \includegraphics[]{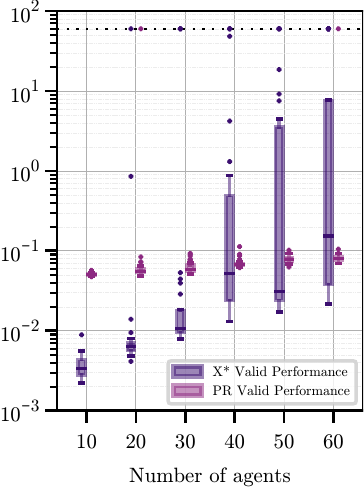}
  \caption{X* vs PR 10\%}
\end{subfigure}
\caption{Box plots of time to valid path
  for X* vs PR. For each agents count, 30 trials are run, each with a 1 minute timeout, with
  each trial run on a randomly generated $100 \times 100$
  four-connected grid. The percentage of states blocked is listed in each caption.}
\figlabel{prsolutions}
\end{figure}

\begin{table}[H]
  \footnotesize
\centering
\setlength\tabcolsep{1.5pt}
    \compresscaptions
\begin{tabular}{|l|l|l|l|l|}
\hline
\textbf{Scenario} & \textbf{X*}         & \textbf{CBS}              & \textbf{M*} & \textbf{AFS} \\ \hline
den520d  & \begin{tabular}[c]{@{}l@{}}0.0026\\ 0.0027\end{tabular}  & \begin{tabular}[c]{@{}l@{}}--\\ 0.0024\end{tabular} & \begin{tabular}[c]{@{}l@{}}--\\  4.2496\end{tabular} & \begin{tabular}[c]{@{}l@{}}12.0885\\ 12.0885\end{tabular} \\ \hline
brc202d & \begin{tabular}[c]{@{}l@{}} 0.0038\\  0.0050\end{tabular} & \begin{tabular}[c]{@{}l@{}}--\\  0.0037\end{tabular} & \begin{tabular}[c]{@{}l@{}}--\\  4.6910\end{tabular} & \begin{tabular}[c]{@{}l@{}} 8.1243 \\  8.1310\end{tabular} \\ \hline
lak303d  & \begin{tabular}[c]{@{}l@{}} 0.0023\\  0.0052\end{tabular} & \begin{tabular}[c]{@{}l@{}}--\\ 0.0023\end{tabular} & \begin{tabular}[c]{@{}l@{}}--\\ 1.5907\end{tabular} & \begin{tabular}[c]{@{}l@{}} 2.6335\\  2.6335\end{tabular} \\ \hline
ht\_mansion\_n  & \begin{tabular}[c]{@{}l@{}} 0.0021\\  0.0035\end{tabular} & \begin{tabular}[c]{@{}l@{}}--\\ 0.0017\end{tabular} & \begin{tabular}[c]{@{}l@{}}--\\  0.7301\end{tabular} & \begin{tabular}[c]{@{}l@{}}0.7354\\  0.7357\end{tabular} \\ \hline
ost003d & \begin{tabular}[c]{@{}l@{}} 0.0022\\  0.0037\end{tabular} & \begin{tabular}[c]{@{}l@{}}--\\  0.0018\end{tabular} & \begin{tabular}[c]{@{}l@{}}--\\  1.4160\end{tabular} & \begin{tabular}[c]{@{}l@{}} 2.1470\\  2.1470\end{tabular}  \\ \hline
w\_woundedcoast & \begin{tabular}[c]{@{}l@{}}0.0104\\  0.0180\end{tabular} & \begin{tabular}[c]{@{}l@{}}--\\  0.0173\end{tabular} & \begin{tabular}[c]{@{}l@{}}--\\ 3.1426\end{tabular} & \begin{tabular}[c]{@{}l@{}} 1.9448\\  1.9466\end{tabular} \\ \hline
\end{tabular}
\caption{X*, CBS, AFS, and M* run on various standard benchmarks
  for 50 agents on all 25 provided random instances with a timeout of 300
  seconds. Median time to valid path is
  reported in the first row and time to optimal path is reported in the
  second row in seconds.}
\tablelabel{CommonBenchmarks}
\end{table}

\subsection{Comparison for time to Optimal Path}\subsectionlabel{timetooptimalsolution}

\begin{figure}[htbp!]
  \compresscaptions
\centering
\begin{subfigure}[b]{0.3\textwidth}
  \centering
  \includegraphics[]{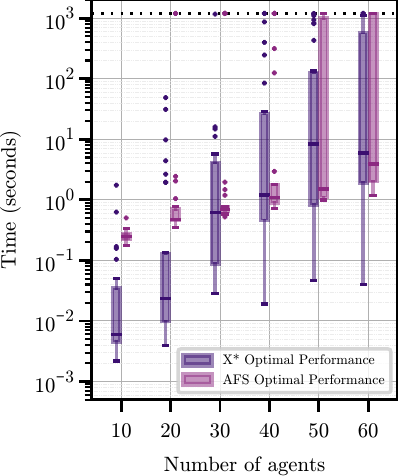}
  \caption{X* vs AFS 1\%}
\end{subfigure}
\hfill
  \begin{subfigure}[b]{0.3\textwidth}
  \centering
  \includegraphics[]{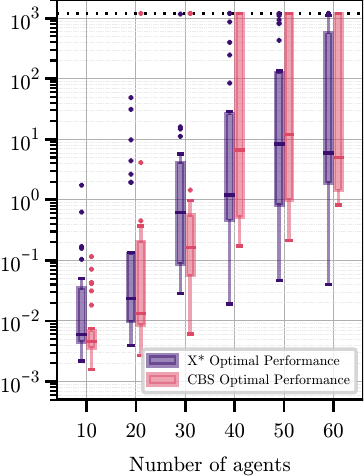}
  \caption{X* vs CBS 1\%}
\end{subfigure}
\hfill
\begin{subfigure}[b]{0.3\textwidth}
  \centering
  \includegraphics[]{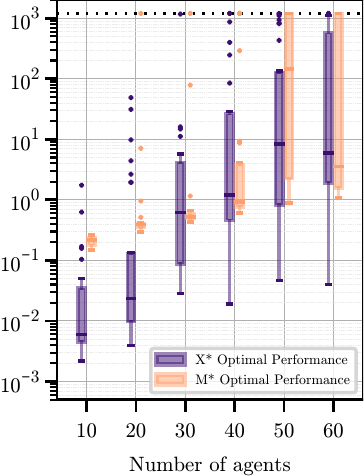}
  \caption{X* vs M* 1\%}
\end{subfigure}

\begin{subfigure}[b]{0.3\textwidth}
  \centering
  \includegraphics[]{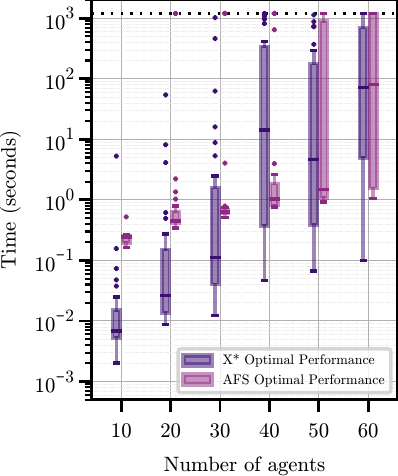}
  \caption{X* vs AFS 5\%}
\end{subfigure}
\hfill
  \begin{subfigure}[b]{0.3\textwidth}
  \centering
  \includegraphics[]{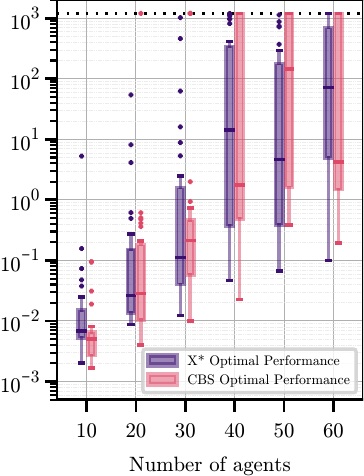}
  \caption{X* vs CBS 5\%}
\end{subfigure}
\hfill
\begin{subfigure}[b]{0.3\textwidth}
  \centering
  \includegraphics[]{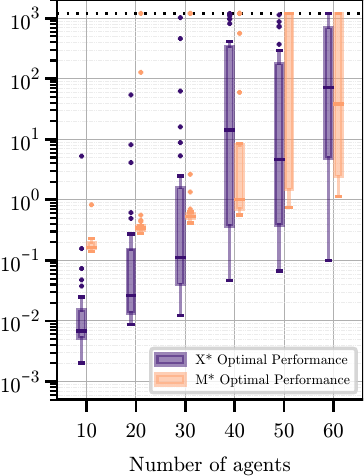}
  \caption{X* vs M* 5\%}
\end{subfigure}

\begin{subfigure}[b]{0.3\textwidth}
  \centering
  \includegraphics[]{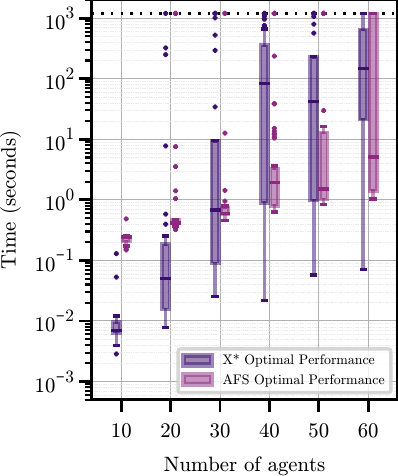}
  \caption{X* vs AFS 10\%}
\end{subfigure}
\hfill
  \begin{subfigure}[b]{0.3\textwidth}
  \centering
  \includegraphics[]{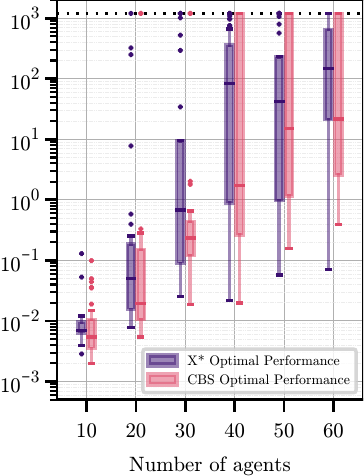}
  \caption{X* vs CBS 10\%}
\end{subfigure}
\hfill
\begin{subfigure}[b]{0.3\textwidth}
  \centering
  \includegraphics[]{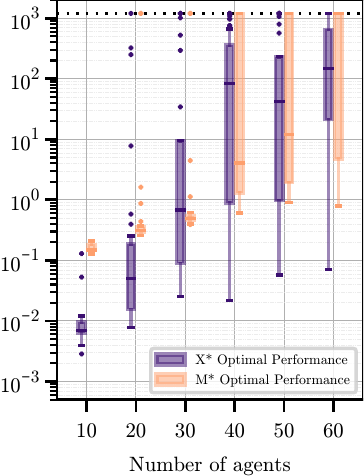}
  \caption{X* vs M* 10\%}
\end{subfigure}
\caption{Box plots of time to optimal path
  for X*, AFS, CBS, and M* on a log scale. For each agents count, 30 trials are run, each with a 20 minute timeout, with
  each trial run on a randomly generated $100 \times 100$
  four-connected grid. The percentage of states blocked is listed in each caption.}
\figlabel{optimalsolutiongrid}
\end{figure}

In order to evaluate the performance of X* compared to state-of-the-art \mapf{} solvers for time to optimal path generation in sparse domains, we run X*, AFS, CBS, and M* with varied numbers of agents on randomly generated $100\times 100$ grids with 1\%, 5\%, and 10\% of the states blocked (\figref{optimalsolutiongrid}) and on standard benchmark domains for fixed number of agents (\tableref{CommonBenchmarks}, second rows).

\figref{optimalsolutiongrid} demonstrates that in random domains, X* is competitive with state-of-the-art \mapf{} planners in time to an optimal path. Like with time to valid path, X* is most competitive when the domains are sparser, i.e.\ lower numbers of agents or fewer blocked states. Against AFS and M*, for small numbers of agents X* exhibits a significantly lower mean and lower quartile runtime; for larger numbers of agents, X* exhibits similar or higher means and but significantly faster lower quartile times. Against CBS, X* has either higher or similar means with heavily overlapping interquartile ranges and lower quartiles. As discussed in \subsectionref{XStarDominantComponenets} the variance in X*'s optimal path generation time can be attributed to the variance in the number of agents involved in any window search; as X* repeatedly grows windows the likelihood window merges thus requiring higher dimensional searches increases, contributing to the large spread in runtimes.

\tableref{CommonBenchmarks} show that X*'s is significantly faster than AFS and M* for time to optimal path generation with over a two order of magnitude faster time, while CBS and X* are highly competitive; this is a reflection of the high sparsity of the benchmark domains and the initial overhead of AFS and M*.

\subsection{X* Versus Baselines}\subsectionlabel{XStarVsBaselines}

X* operates by restricting the initial
repair search space, quickly finding a repair in this restricted search space to produce a valid global path, then relaxing the restriction 
and repeating the process until an optimal global path is found. While this
approach provides \ampp{}'s anytime property, it also incurs
computational overhead, even in planners like X* which perform reuse between
repair searches.

To demonstrate this overhead, we run X*, NWA* and A* on a $20
\times 20$ four-connected grid scenario with an agent starting on the center of each edge and
with a goal on the center of the opposite edge, thereby inducing
a four agent collision in the center of the scenario. While A* will directly solve
for an optimal path, X* and NWA* will quickly produce a valid global path, multiple intermediary global paths, and terminate with a provably optimal global path. 

The runtime results are presented in \tableref{BaselineTable}, with 95\% confidence
intervals over 30 trials. Due to the nearly identical structure of their
 initial path generation, NWA* and X* have nearly identical performance for time to first
path, outperforming A*'s time to its first path by over an order of
magnitude. Due to the window
overhead, X* takes approximately 1.5 times longer than A* to produce an optimal global
path, having finished 5 of the needed 9 window expansions when A*
terminates, and NWA* takes approximately 6x longer than A* to produce an optimal global path due to a lack of
search re-use, having finished 3 of the needed 9 window expansions when A* terminates.
This result demonstrates the efficacy of X*'s search reuse techniques in improving its optimal path generation performance and demonstrates to practitioners that, while X* and NWA* have the same first path runtime, X* strictly dominates NWA* in time to optimal path.


\vspace{-0.25em}
\begin{table}[H]\footnotesize
  \centering
\begin{tabular}{|l|r|r|r|r|}
\hline
  \textit{Planner} & $\frac{\textup{\textit{Valid Path Runtime}}}{\textup{\textit{med.\ A* runtime}}}$ & $\frac{\textup{\textit{Optimal Path Runtime}}}{\textup{\textit{med.\ A* runtime}}}$ & \textit{curr.\ iter.}  & \textit{total iter.} \\ \hline
X*               &  6.32\% &  175.18\%& 6 & 9 \\ \hline
NWA*             &  6.28\% &  547.20\% & 4 & 9 \\ \hline
A*             &  100.00\% &  100.00\% & -- & --\\ \hline
\end{tabular}
  \caption{X*, NWA*, and A* run on a $20\times20$ grid with an agent starting on the center of each edge and with a goal on the center of the opposite edge to demonstrate the overhead of \ampp{}-style window growth compared to A*. Each result is reported as a percentage of the total A* runtime.  Column \textit{curr.\ iter.} represents the {\recampp{}} iteration the given planner was on when A* terminated with the optimal solution and column \textit{total iter.} represents the total {\recampp{}} iterations needed to generate an optimal solution.}
\tablelabel{BaselineTable}
\end{table}

\begin{figure}[htbp!]
  \compresscaptions
\centering
\begin{subfigure}[b]{0.49\textwidth}
  \centering
    \begin{subfigure}[b]{\textwidth}
  \centering
  \includegraphics[]{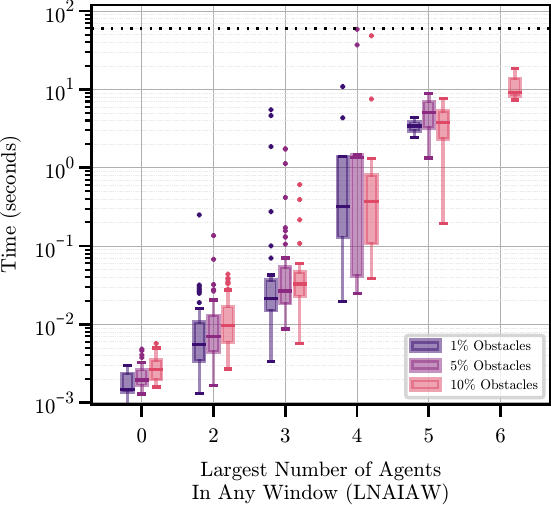}
\end{subfigure}
\begin{subfigure}[b]{\textwidth}
  \centering
  \includegraphics{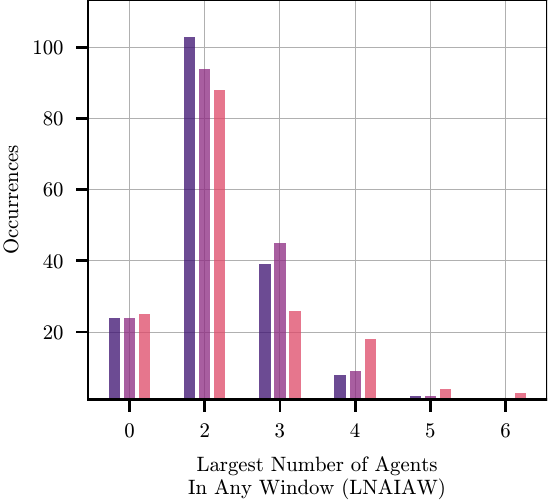}
\end{subfigure}
\caption{Valid Path}
\figlabel{AgentsWindowVsFirstRuntime}
\figlabel{OccurrencesAgentsWindowsFirst}
\end{subfigure}
\hfill
  \begin{subfigure}[b]{0.49\textwidth}
\begin{subfigure}[b]{\textwidth}
  \centering
  \includegraphics[]{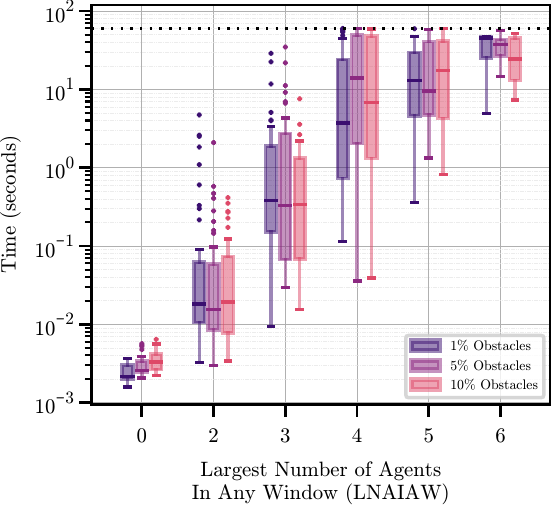}
\end{subfigure}
\begin{subfigure}[b]{\textwidth}
  \centering
  \includegraphics{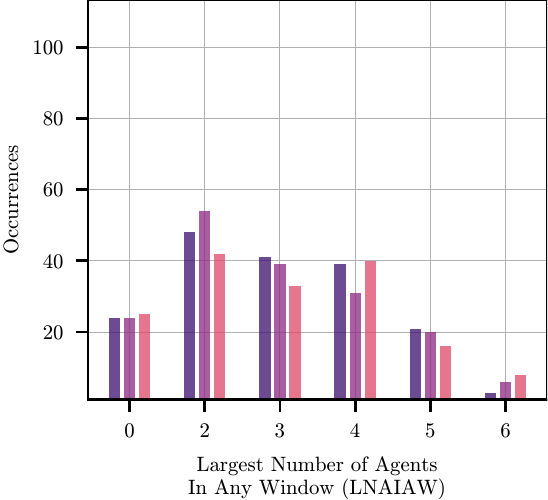}
\end{subfigure}
\caption{Optimal Path}
\figlabel{AgentsWindowVsOptimalRuntime}
\figlabel{OccurrencesAgentsWindowsOptimal}
\end{subfigure}
\caption{Time to valid  and optimal paths
  for X* on a log scale vs Largest Number of Agents In Any Window (\lnaiaw{}), along with the frequency of each \lnaiaw{}. Run
  across 20 to 60 agents in steps of 10, each of 30 trials, on $100 \times 100$ four-connected grids
  with $1\%$, $5\%$, and $10\%$ obstacle density with a timeout of 60 seconds.}
  \figlabel{AgentsWindowVs}
\end{figure}

\subsection{X* Components That Dominate Runtime}\subsectionlabel{XStarDominantComponenets}

In order to optimize X*, be it from an implementation
perspective or a theoretical one, is important to understand which components dominate its
runtime. X*'s runtime is dominated by \callsmall{\planin}{} and
\callsmall{\growandreplan}{}, where the window searches with the highest number of
agents dominate both time to valid path (\figref{AgentsWindowVsFirstRuntime}) and time
to optimal path (\figref{AgentsWindowVsOptimalRuntime}). Fortunately, for
random domains with various agent counts, as the magnitude of the Largest Number of Agents In Any Window (\lnaiaw{})
grows linearly, the number of occurrences of such a window decreases
exponentially for valid path generation (\figref{OccurrencesAgentsWindowsFirst})
and linearly for optimal path generation (\figref{OccurrencesAgentsWindowsOptimal}).
This finding also provides an opportunity for practitioners to build an X*-based composite \ampp{} solver that falls back on another \mapf{} solver when a high dimensional window is detected, preventing X* from performing a potentially expensive search. 

\subsection{X* Runtime Versus Sparsity of
  Domain}\subsectionlabel{XStarRuntimeVsSparsity}

X* is designed to exploit \emph{sparsity} of
agent-agent collisions in order to quickly develop a
suboptimal but valid path as well as produce an optimal global path. 
First, to demonstrate that X* does exploit available sparsity in practice, we look at X*'s success at keeping the number of agents involved in each window low, measured by the magnitude of the Largest Number of Agents In Any Window (\lnaiaw{}). \figref{AgentsWindowVs} demonstrates that as the obstacle density of the domain increases, and thus sparsity decreases, the magnitude of \lnaiaw{} increases; this is especially clear in time to valid path (\figref{AgentsWindowVsFirstRuntime}), where there is a clear increasing trend in the distribution of \lnaiaw{} from 1\% occupied to 10\% occupied grids, but a similar trend exists in time to optimal path (\figref{AgentsWindowVsOptimalRuntime}). These trends are the result of the fact that in domains with relatively high sparsity, e.g.\ the 1\% occupied grids, X* is able to cleanly separate collisions from one another, while in less sparse domains, e.g.\ the 10\% occupied grids, X* cannot separate collisions as well and must form windows with more agents.

\begin{figure}[htbp!]
  \compresscaptions
\centering
\begin{subfigure}[b]{0.3\textwidth}
  \centering
  \includegraphics[]{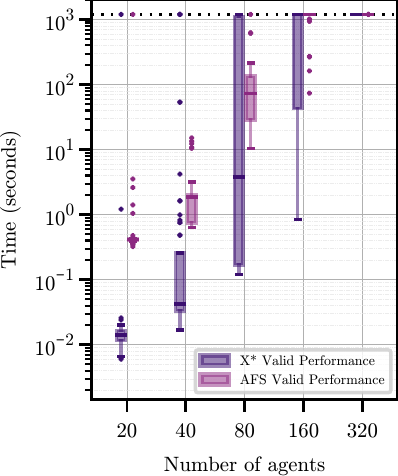}
  \caption{X* vs AFS}
\end{subfigure}
\hfill
  \begin{subfigure}[b]{0.3\textwidth}
  \centering
  \includegraphics[]{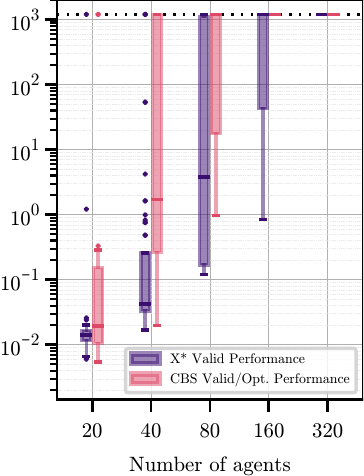}
  \caption{X* vs CBS}
\end{subfigure}
\hfill 
\begin{subfigure}[b]{0.3\textwidth}
  \centering
  \includegraphics[]{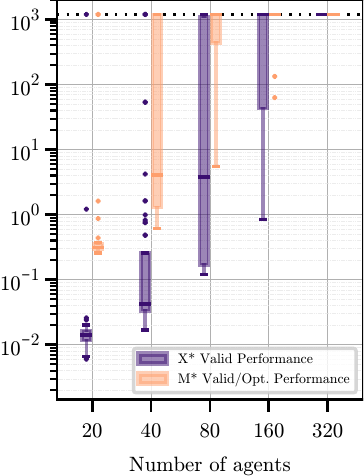}
  \caption{X* vs M*}
\end{subfigure}
\caption{Time to valid path 
  for X* vs AFS, CBS, and M* on a log scale. For each agent count, 30 trials are run, each with a 20 minute timeout, with
  each trial run on with a constant $\frac{\textup{grid area}}{\textup{agent
      count}}$ ratio of 500 with a $10\%$ obstacle density.}
\figlabel{densityfirsts}
\end{figure}

Second, as a result of the fact that X* exploits sparsity, it is expected that X* will scale well when the number of
agents in a domain increases but the level of sparsity stays the same. To
validate this expectation, we run X* on varying sized four-connected grids with
a $10\%$ obstacle density and constant $\frac{\textup{grid area}}{\textup{agent
    count}}$ ratio of 500 in an attempt to maintain similar levels of domain sparsity. We also run CBS, AFS, and M* on the same
domains to provide a frame of reference. Time to valid global path is presented in
\figref{densityfirsts} and time to optimal global path is presented in
\figref{densityoptimals}.


 For time to valid path, X*'s median time is consistently faster than any other planner, its lower quartile is consistently two orders of magnitude faster than AFS or M* and it scales better than any other planner; with the exception of a few instances solved by AFS and M*, X* was the only planner able to produce paths for the 160 agent case, in some cases producing paths in under a second;  X*'s superior performance against these other planners is due to its ability to exploit domain sparsity to greater effect.


\begin{figure}[]
    \compresscaptions
  \centering

  \begin{subfigure}[b]{0.3\textwidth}
    \centering
    \includegraphics[]{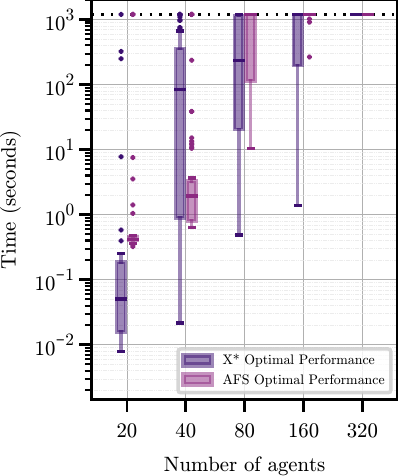}
    \caption{X* vs AFS}
  \end{subfigure}
  \hfill
    \begin{subfigure}[b]{0.3\textwidth}
    \centering
    \includegraphics[]{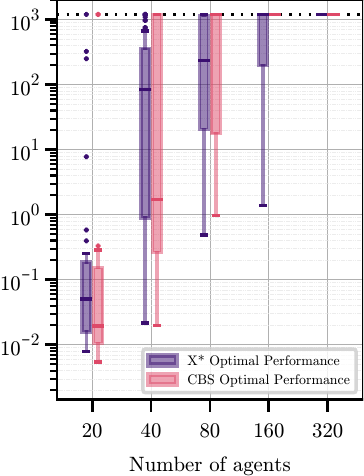}
    \caption{X* vs CBS}
  \end{subfigure}
  \hfill 
  \begin{subfigure}[b]{0.3\textwidth}
    \centering
    \includegraphics[]{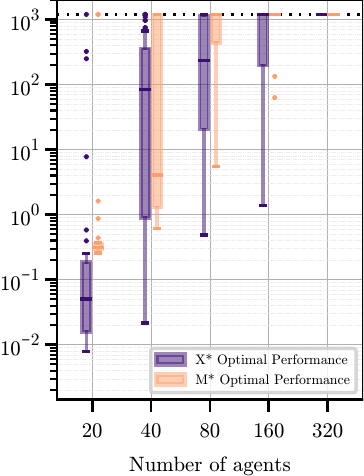}
    \caption{X* vs M*}
  \end{subfigure}
  \caption{Time to optimal path 
    for X* vs AFS, CBS, and M* on a log scale. For each agent count, 30 trials are run, each with a 20 minute timeout, with
    each trial run on with a constant $\frac{\textup{grid area}}{\textup{agent
        count}}$ ratio of 500 with a $10\%$ obstacle density.}
  \figlabel{densityoptimals}
\end{figure}

For time to optimal path, X* has a higher median runtime than the other
planners for lower numbers of agents; however, for 80 agents, X*'s median runtime is
below the timeout threshold while all other planners medians are at the timeout
threshold and, with the exception of a few instances solved by AFS and M*, X* is the only planner able to generate optimal paths for 160 agents. 

Together, these findings suggest that, compared to state-of-the-art algorithms,
X*'s approach scales well to large numbers of agents across domains with similar levels of sparsity.

\subsection{Suboptimality Bounds on Intermediary Paths}\subsectionlabel{suboptimalitybounds}

For the $\epsilon$-suboptimal intermediary solutions of an anytime planner to be useful in practice, the $\epsilon$ bound must be reasonably tight. In order to characterize X*'s $\epsilon$ bound in practice, we ran X* for 30 trials on a $100 \times 100$ random grid with 30 agents and varied obstacle density. The results for the first 20 X* iterations (recursive invocations of \callsmall{\recampp}), shown in \figref{suboptimalitybounds}, were collected from the same experiments shown in \figref{validsolutiongrid} and \figref{optimalsolutiongrid}.

\begin{figure}[htbp!]
  \compresscaptions
\centering
\begin{subfigure}[b]{0.33\textwidth}
  \centering
  \includegraphics[]{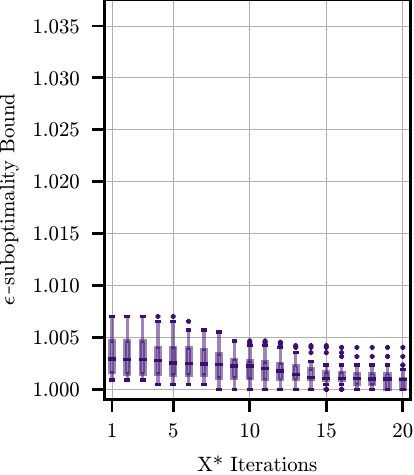}
  \caption{1\% Obstacle Density}
\end{subfigure}
\hfill
  \begin{subfigure}[b]{0.3\textwidth}
  \centering
  \includegraphics[]{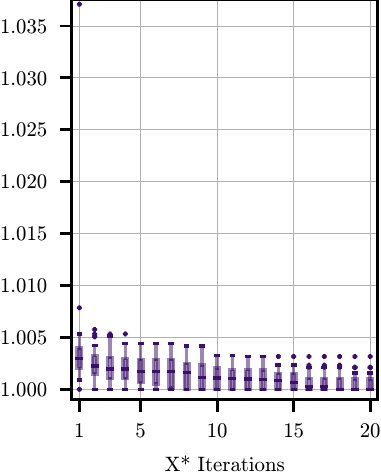}
  \caption{5\% Obstacle Density}
\end{subfigure}
\hfill 
\begin{subfigure}[b]{0.3\textwidth}
  \centering
  \includegraphics[]{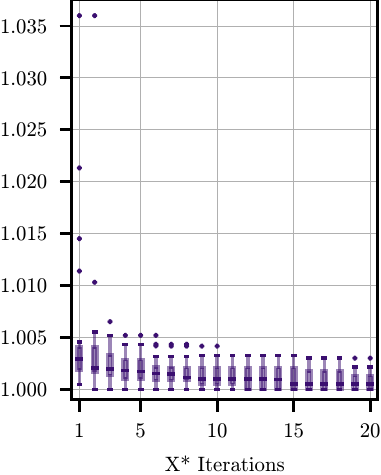}
  \caption{10\% Obstacle Density}
\end{subfigure}
\caption{$\epsilon$-suboptimality bounds for the first 20 iterations of 30 trials of X* on $100 \times 100$ random grids for 30 agents. The 3 trials that failed to produce any path in 10\% Obstacle Density were not recorded. The trials that terminated in fewer than 20 iterations had their last bound duplicated for the remaining iterations.}
\figlabel{suboptimalitybounds}
\end{figure}

These results demonstrate that, in practice, X*'s first valid path cost is almost always within 0.5\% of the optimal path and outliers are quickly improved upon within a few additional iterations of X*. For practitioners, these results indicate that X*'s first path is often of sufficient quality and, if not, a few additional iterations of \callsmall{\recampp} should be sufficient to bring the path quality within a tight quality bound.

\subsection{X* Window Selection Impact on Runtime}\subsectionlabel{XStarWindowSelection}

\begin{figure}[htbp!]
    \compresscaptions
  \centering
  \begin{subfigure}[b]{0.49\textwidth}
    \centering
    \includegraphics[]{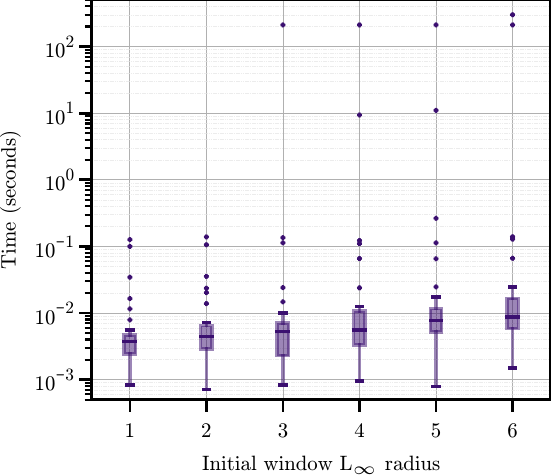}
    \caption{Valid Path}
    \figlabel{RadiusFirstSolution}
  \end{subfigure}
  \hfill
  \begin{subfigure}[b]{0.49\textwidth}
    \centering
    \includegraphics[]{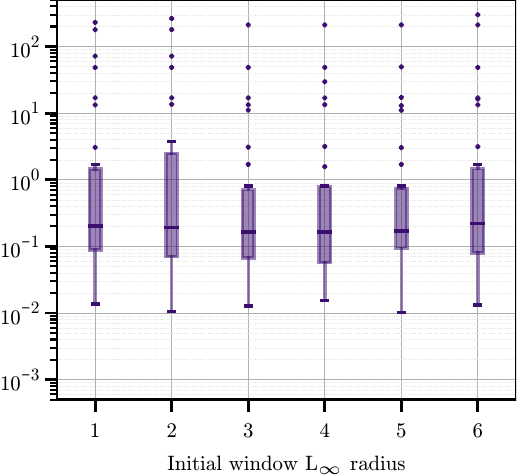}
    \caption{Optimal Path}
    \figlabel{RadiusOptimalSolution}
  \end{subfigure}
\caption{Time to valid path and time to optimal path vs initial window radius 
    for X* on a log scale. Run
    across 30 trials of 30 agents on $100 \times 100$ four-connected grids
    with $5\%$ obstacle density.}
  \label{fig:three graphs}
\end{figure}

As shown in \subsectionref{XStarDominantComponenets}, window dimensionality 
dominates runtime. As such, selecting the proper initial window size to repair a
search in order to minimize window merges is an important factor in X*'s valid
path generation performance. \figref{RadiusFirstSolution} shows the impact
of the initial window radius parameter on X*'s time to valid path;
unsurprisingly, smaller window radii more quickly produce a valid global path due to a decreased likelihood of
requiring window merges.

However,
smaller window radii can \emph{increase} time to optimal path
in some cases. Shown in \figref{RadiusOptimalSolution}, an initial window radius
of 1 or 2 result interval bounds that are roughly 5x higher than the bounds produced by initial window radii of  3, 4,
and 5, with similar performance differences even in outliers. The root cause of this performance degradation is the
expansion of states during a small window search which would not be expanded
by a fresh search in a larger window, such as depicted in the large dark blue area of \figref{bookkeepingexamplesreexpansion}. As
such, these unnecessary expansions earlier in X*'s search will add states to
$\openset$ to be expanded which would never be considered by a search that
initially had a larger window. The exact
radius values for which performance degrades changes across scenarios
as a consequence of the structure of the domain, making this analysis important
for practitioners who care about time to optimal path.

\section{Future Work}\sectionlabel{FutureWork}

X* uses standard A* to perform optimal window searches; if a fast optimal MAPF solver such as CBS or an anytime MAPF solver such as AFS were used to admit suboptimal repairs inside a window,
and this search tree could be grown using X*-style reuse, this approach may
produce a WAMPF planner faster than X*. This investigation would also
lend itself to exploring $\epsilon$-suboptimal \ampp{}.

In addition, there is room for further exploration of window size and shape; in this work we used rectangular windows for NWA* and X* because they are simple to reason about and performed better in our initial experimentation than rasterized spheres, but there may be other shapes that are better suited to \ampp{}.

Finally, we believe that further investigation into quantifying sparsity of
\mapf{} domains would provide great insight into the fundamental nature of
\mapf{} and potentially allow for the development of an ensemble MAPF solver that
switches techniques based on individual problem structure.

\section{Acknowledgements}

This work is supported in part by AFRL and DARPA under agreement \#FA8750-16-2-0042, and NSF grants IIS-1724101 and IIS-1954778. We would also like to thank Liron Cohen for his implementation of AFS, Wolfgang Hoenig for his implementation of CBS, Glen Wagner for his implementation of M*, and Ilja Ivanashev for his implementation of PR. Finally, we would like to thank the anonymous reviewers for their helpful feedback that improved this work.

\section{Bibliography}

\bibliography{bibliography}

\appendix

\setcounter{theorem}{0}
\setcounter{myproof}{0}

\section{\ampp{} Proofs}\appendixlabel{amppproofs}

\begin{theorem}\theoremlabel{AMPPValidSolutionImmediately}
  If we assume \Call{\planin}{} and \Call{\growandreplan}{} produce optimal paths in $\window$, a valid path exists, then \ampp{} will produce a valid path after a single invocation of \Call{\recampp}{}.
\end{theorem}

\begin{myproof}\prooflabel{AMPPValidSolutionImmediately}
  This is a special case of Case \ref{AlreadyOptimalCase} or Case
  \ref{NotAlreadyOptimalCase} in \proofref{AMPPOptimal}; as shown, either
  $\pth{}$ generated on \lineref{AMPPPlanIndependently} is optimal, in which
  case \ampp{} terminates with $\pth{}$ as its path, or $\pth$ will be
  repaired to generate a valid path.
\qed
\end{myproof}

\begin{theorem}\theoremlabel{AMPPOptimal}
  If we assume:
  \begin{enumerate}
    \item A valid path exists.
    \item \Call{\planin}{} and \Call{\growandreplan}{} produce optimal repairs in their given windows.
    \item \Call{\shouldquit}{$\windowk$} does not discard 
    a window $\windowk$ with an associated agent set $\agentset$ until $\startk = \filterpath{\start}{\agentset}$, $\goalk= \filterpath{\goal}{\agentset}$, and the repair is unimpeded by $\windowk$'s restrictions on the state space.
  \end{enumerate}
   Then, given sufficient time \ampp{} will produce a minimal cost path.
\end{theorem}

\begin{myproof}{}\prooflabel{AMPPOptimal}
\begin{lemma}[Optimal merged paths are optimal]\lemmalabel{CombineOptimalPaths}
Given two paths, $\pth$ for agent set $\agentset$ and $\pth'$ for agent set 
$\agentset'$, where $\pth$ and $\pth'$ are optimal,  $\agentset \cap \agentset' = \emptyset$, and
$\pth$ and $\pth'$ do not collide with each other, then if
$\pth$ and $\pth'$ are joined to produce $\pth''$, it
follows that $\pthcost{\pth''} =
\pthcost{\pth} + \pthcost{\pth'}$, and thus $\pth''$ is optimal.

Proof by contradiction: Consider a case where $\pth''$ constructed via the
method above is not optimal. That would imply that there exists another,
optimal path with the same $\start$ and $\goal$, $\pth'''$, such that:

\begin{align*}
  \pthcost{\pth'''} &= \pthcost{\filterpath{\pth'''}{\agentset}} +
                        \pthcost{\filterpath{\pth'''}{\agentset'}}\\
                      &< \pthcost{\pth''}\\
                      &= \pthcost{\filterpath{\pth''}{\agentset}} +
                       \pthcost{\filterpath{\pth''}{\agentset'}}\\
                    & = \pthcost{\pth} +  \pthcost{\pth'}\\
                     &\implies\\
  \pthcost{\filterpath{\pth'''}{\agentset}} < \pthcost{\pth} &\lor 
  \pthcost{\filterpath{\pth'''}{\agentset'}} < \pthcost{\pth'}
\end{align*}

which implies that $\pth$ or $\pth'$ are suboptimal, which violates the
assumption that $\pth$ and $\pth'$ are optimal.
\qed
\end{lemma}

\begin{lemma}[Unrestricted window searches produce optimal paths]\lemmalabel{OptimalRepairFound}
Given a joint path $\pth$ and 
window $\window$ with an associated agent set $\agentset$ is used to repair
$\filterpath{\pth}{\agentset}$, if $\window$ contains $\start$ and $\goal$, the global start and goal for agents $\agentset$, and a repair within $\window$ has just been performed on $\filterpath{\pth}{\agentset}$  in which $\window$ did not constrain
the search between $\start$ and $\goal$, then $\filterpath{\pth}{\agentset}$ is globally optimal for agent set $\agentset$.

We know from the
definition of a window that if $\start$ and $\goal$
associated with $\filterpath{\pth}{\agentset}$ are in $\window$, then they are the $\start$ and
$\goal$ used by $\window$. Thus, we know that as the given repair strategy
produces an optimal solution in $\window$ between $\window$'s $\start$ and $\goal$, 
$\window$'s $\start$ and $\goal$ are $\start$ and $\goal$ of
$\filterpath{\pth}{\agentset}$, and the search was not restricted by $\window$,
then this path would be optimal even for an arbitrarily large $\window$, and
thus $\filterpath{\pth}{\agentset}$ is globally optimal for agent set $\agentset$.
\qed
\end{lemma}

\begin{lemma}[\Call{\recampp}{} always grows all windows]\lemmalabel{WindowsGrow}
  Given a set of windows $W$, all $\window \in W$ will be enlarged by
  $\Call{\recampp}{}$ to encompass more states.

At the start of each iteration of $\Call{\recampp}{}$,
$\Call{\growandreplan}{}$ will be invoked $\forall \window \in W$
(\linesref{AMPPGrowAndReplanStart}{AMPPGrowAndReplan}), by definition causing
all windows to be grown, thereby upholding the claim. Some of these windows may be merged with
existing windows (\lineref{AMPPGrowAndReplanEnd}), resulting in a larger,
merged windows (\lineref{AMPPUnionWindows}), thereby upholding the claim. Some of
these windows may be merged with newly created windows, resulting in larger,
merged windows (\lineref{AMPPEndWhileExistsCollision}), thereby upholding the claim.
\qed 
\end{lemma}

\noindent When \Call{\recampp}{} is called, we know a given path is either:

\begin{enumerate}{}
\item Valid and globally optimal, with $W = \emptyset$\label{AlreadyOptimalCase}
\item Invalid and at or below cost of globally optimal solution, with $W = \emptyset$\label{NotAlreadyOptimalCase}
\item Valid and potentially globally suboptimal, with windows surrounding locally
  optimal repairs, i.e.\ $W \neq \emptyset$\label{ConvergeCase}
\end{enumerate}

\noindent We do an analysis of \Call{\recampp}{} in these three cases:

\begin{enumerate}{}
\item We invoke \Call{\recampp}{} with a valid and globally optimal solution and $W =
  \emptyset$.
  \linesref{AMPPGrowAndReplanStart}{AMPPGrowAndReplanEnd} are skipped, as $W = \emptyset$.
\linesref{AMPPWhileExistsCollision}{AMPPEndWhileExistsCollision} are skipped, as 
no collisions exist. \linesref{AMPPShouldQuitLoop}{AMPPShouldQuit} are
skipped, as $W = \emptyset$. Finally, $W = \emptyset$, so $\left( \pi, 1 \right)$ is returned with
$\pth$ unmodified (\lineref{AMPPReportDone}) and thus \Call{\recampp}{}
returns $\pth$, having proved it's a globally optimal solution.

\item We invoke \Call{\recampp}{} with an invalid solution at or below joint optimal
  cost and $W = \emptyset$.
  \linesref{AMPPGrowAndReplanStart}{AMPPGrowAndReplanEnd} are skipped, as $W =
  \emptyset$. \linesref{AMPPWhileExistsCollision}{AMPPEndWhileExistsCollision} 
  create windows and locally repair each collision as they occur along $\pth$,
  merging windows if they overlap. When
  \linesref{AMPPWhileExistsCollision}{AMPPEndWhileExistsCollision}  are
  complete, $\pth$ is a valid but potentially globally suboptimal solution.
  If \linesref{AMPPShouldQuitLoop}{AMPPShouldQuit} can prove that all windows
  produces local repairs that are globally optimal, then \Call{\recampp}{}
  returns $\pth$, having proven it's a  globally optimal solution. Otherwise,
  \Call{\recampp}{} has produced a valid and potentially globally suboptimal
  solution with windows surrounding locally optimal repairs, the scenario handled by
  Case~\ref{ConvergeCase}.
  
\item We invoke \Call{\recampp}{} with a valid and
  potentially globally suboptimal path $\pth$ with windows surrounding locally optimal
  repairs. We know from \lemmaref{WindowsGrow} that these windows will continue
  to grow with each recursive invocation of \Call{\recampp}{}, any overlapping
  windows will be merged together
  (\linesref{AMPPCheckInterfere}{AMPPGrowAndReplanEnd}), and any new collisions
  induced by repairs will be encapsulated by a new window and merged with any
  overlapping existing windows
  (\linesref{AMPPWhileExistsCollision}{AMPPEndWhileExistsCollision}). Thus, we
  know in a finite number of recursive invocations of \Call{\recampp}{}, every
  window $\window$, associated with an agent set $\agentset$,  will eventually
  contain $\start$ and $\goal$ associated with $\filterpath{\pth}{\agentset}$
  such that the window based repair between $\start$ and $\goal$ is not
  constrained by $\window$. Thus, we know from \lemmaref{OptimalRepairFound}
  that the globally optimal path for $\agentset$ from $\start$ to $\goal$ has
  been proven to be found, and thus $\window$ can be removed from $W$ by
  \Call{\shouldquit}{} (\linesref{AMPPShouldQuitLoop}{AMPPShouldQuit}). Thus,
  after a finite number of iterations, \Call{\recampp}{} will terminate and from
  \lemmaref{CombineOptimalPaths} we know that the globally optimal solution has been found.

\end{enumerate}

\noindent We know that \Call{\recampp}{} will only be invoked in the three cases:

\begin{enumerate}{}
\item $\pth$ is composed of individually planned, optimal paths
  (\lineref{AMPPPlanIndependently}), and it is collision free. $W = \emptyset$
  (\lineref{AMPPWEmpty}), and so it qualifies for Case \ref{AlreadyOptimalCase}.
  Case \ref{AlreadyOptimalCase} always terminates after a single invocation of
  \Call{\recampp}{}, and $\pth$ has been proved to be optimal.

\item $\pth$ is composed of individually planned, optimal paths
  (\lineref{AMPPPlanIndependently}), and it is \emph{not} collision free. $W = \emptyset$
  (\lineref{AMPPWEmpty}), and so it qualifies for Case
  \ref{NotAlreadyOptimalCase}. Case \ref{NotAlreadyOptimalCase} either
  terminates after a single invocation of \Call{\recampp}{}, and $\pth$ has
  been proved to be optimal, or it invokes \Call{\recampp}{} in Case
  \ref{ConvergeCase}.

\item $\pth$ is in the process of being repaired, making it potentially
  globally suboptimal, and it has an associated window set $W \neq \emptyset$.
  Case \ref{ConvergeCase} either terminates and $\pth$ has
  been proved to be optimal, or it again invokes Case \ref{ConvergeCase}.    \qed
\end{enumerate}
\end{myproof}

\setcounter{theorem}{0}
\setcounter{myproof}{0}

\section{X* Proofs}\appendixlabel{xstar}

\begin{theorem}\theoremlabel{Stage2OptCollisionFree}
  During \stagetwo{}, the path between $\startkpone$ and $\startk$ extracted from the full joint path $\pth$ is collision-free.
\end{theorem}
\begin{myproof}
  We know that after a single iteration of \callsmall{\recampp}, the full joint path $\pth$ will be collision free (\appendixref{amppproofs}, \theoremref{AMPPValidSolutionImmediately}). Additionally, we know by construction that \callsmall{\growandreplan} will not be invoked until after the first iteration of \callsmall{\recampp}, and it will be invoked on the windows created during the initial iteration of \callsmall{\recampp}. All subsequent changes to $\pth$ via window growth or merging are improvements; X*'s \callsmall{\planin} and \callsmall{\growandreplan} ensure that these improvements do not create collisions in regions of $\pth$ not encompassed by existing windows by ensuring that the timing of agent movements in regions not encompassed by existing windows remains unchanged. To do this, \callsmall{\growandreplan} and \callsmall{\planin} ensure that, after the first iteration of \callsmall{\recampp}, repairs within the given window are padded as necessary (\subsectionref{XStarImplementations}). Additionally, \ampp{} ensures that if a $\windowkpone$ overlaps with any another windows, the overlapping windows are merged and repaired via \callsmall{\planin} rather than invoking \callsmall{\growandreplan} on $\windowk$ (\algoref{\ampp}, \linesref{AMPPGrowAndReplanStart}{AMPPGrowAndReplanEnd}), thus ensuring that $\startkpone$ cannot be inside another window. As a result, the section of $\pth$ from $\startkpone$ to $\goalkpone$ must be from a region of $\pth$ unencompassed by another window, and regions of $\pth$ unencompassed by another window must be collision free, and thus the section of $\pth$ from $\startkpone$ to $\goalkpone$ is collision-free.
  \qed
\end{myproof}

\end{document}